# Monitoring Thermal Transformations in Hybrid Composites

*György Bánhegyi*[a*], *Zsuzsanna Mátyás-Karácsony*[b], *Éva Fazakas*[b], *Miklós Mohai*[c], *Richárd Bak*[b]

[a]Medicontur Medical Engineering Ltd., Herceghalmi út 1, 2072 Zsámbék, Hungary
[b]Bay Zoltán Nonprofit Ltd. for Applied Research, Kondorfa u. 1, 1116, Budapest, Hungary
[c]Institute of Materials and Environmental Chemistry, Research Centre for Natural Sciences, Hungarian Academy of Sciences, Magyar tudósok körútja 2, 1117 Budapest, Hungary,

**Abstract.** Hybrid composites, mixtures of organic and inorganic materials frequently achieve their final properties after thermal treatment involving partial or complete decomposition and chemical reactions. Three examples are presented to demonstrate the possibilities of monitoring these complex transformations by various analytical techniques. The first one is a polyurea based prepreg system containing both organic and inorganic additives, achieving extreme heat stability after curing followed by a heat treatment. Next example is taken from alumina based ceramic injection molding systems. The effects of matrix polymer and various additives, such as citric acid, UHMWPE (ultrahigh molecular weight polyethylene) and the effect of alumina trihydrate/alumina ratio in the ceramic injection molding feedstock was studied on the processability and porosity. The last example presents a new approach to prepare ODS (oxide dispersed steel) bodies by using metal-organic compounds as additives in metal injection molding feedstocks. Applied characterization methods included thermogravimetry (TGA), differential thermal analysis (DTA), X-ray diffraction (WAXS), X-ray photoelectron spectroscopy (XPS), scanning electron microcopy (SEM) coupled with energy dispersive X-ray analysis (EDX), vibrational spectroscopy (Raman and FTIR) and density measurements.

**Keywords:** Hybrid composite; Heat treatment; Decomposition, Thermal analysis

## 1. Introduction

Composite materials can be defined as micro-heterogeneous materials where usually one or more components are dispersed in a matrix, although there are systems where the dispersed and continuous phases cannot be distinguished from each other (statistical mixtures or interpenetrating networks). An important subclass of composites is that of so-called hybrid composites. Here "hybrid" either means the combination of various types or sizes of fillers in a continuous matrix or the matrix itself is composed of various materials [1, 2]. In some special cases "hybrid" composites mean the intimate (nano-heterogenous) mixture of organic and inorganic components. Certain hybrid composites are exposed to heat treatment, e.g. curing of the matrix material, reaction with the filler, transformation of the metal-organic component into nano-fillers (sol-gel reactions) [3], or even the complete destruction of the organic component (as in powder injection molding, including metal injection molding (MIM) and ceramic injection molding (CIM) [4], ceramic precursor polymer processing [5]). These complex transformations lead to completely new materials or new properties. As the number of combinations is almost infinite, there is a good chance to tailor the final properties of the products obtained (fibers, coatings, powders, bulk materials, freeform articles etc.).

This article briefly presents some examples of hybrid composite materials from our own unpublished research wherein the organic matrix is itself a multi-component material and undergoes either a reaction with the filler followed by a partial or complete thermal degradation or contains special additives to achieve a modification of the final properties of the heat treated final product, which is usually a composite itself. We would like to demonstrate the use of various analytical techniques to monitor and understand these complex thermal transformation processes on three examples from our own research. The experimental background, the results and discussion will be presented separately for three kinds of materials. The first example is an isocyanate based prepreg system intended for high temperature use (chimney lining). Isocyanates are best known as the main ingredients in polyurethanes and polyureas [6], wherein the isocyanate component is reacted with polyols and polyamines respectively. Isocyanates, however, can be combined directly with inorganic compounds, such as aqueous silicate solutions (commonly called

---
[*]Corresponding author, e-mail: gybanhegyi@medicontur.hu

Bánhegyi et al.: Monitoring Thermal Transformations in Hybrid Composites"water glass") to yield resinous materials with very interesting properties [7]. These form the basis of the so-called "silicate resins" [8], although this is a somewhat misleading name, as these systems are not silicate polymers – as opposed to geopolymers [9]. These kinds of isocyanate/water glass resins can be further hybridized with other organic resins, such as epoxy resins [10] or vinylester resins [11]. Such hybrid resins have been patented among others by Polinvent Ltd. [12]. A recently submitted patent [13] describes a prepreg for chimney lining, containing monomeric, oligomeric and polymer methylene diisocyanates (MDI, at least 50 wt%), finely dispersed metal hydroxides such as $Ca(OH)_2$ powder, alumina trihydrate (ATH), optionally other metal oxides or hydroxides, inert inorganic fillers, various flame retardants, plasticizer components and glass or basalt fiber based fabric as reinforcement. The changes in the composition were studied by various methods to understand better the transformation process during heat treatment. Further patents are being prepared on related systems.

The second example is related to ceramic injection molding (CIM) feedstock materials. It is well known that CIM feedstock contains organic binders, which are usually composed of polyolefins, polyolefin copolymers and waxes [14–16]. In addition to the binder composition successful processing of the feedstock to ceramic end-product is influenced by many other factors, such as powder granulometry, rheological properties, details of de-binding (solvent based or purely thermal) and sintering processes [17].

Here we describe and investigate formulations which were used to obtain alumina bodies with controlled porosities. A part of these results have been described in an MSc thesis supervised by one of the authors [18]. Other, closely related results on low pressure injection moldable or slip-casting formulations have been published by us [19]. Some other groups [20, 21] describe methods wherein the ratio of ceramic powder/polymer binder was used to achieve various porosities in CIM processed alumina. Apparently however, pore forming agents or analogs of chemical "blowing agents" used in plastics processing [22] have been rarely mentioned, mostly in relation to ceramic foams [23], although polymer particles (polystyrene, polyurethane foam and hydrogen peroxide) have been used as pore formers in creating relatively high porosity (55–80%) alumina bodies [24] and hot isostatic pressing has also been used to produce porous alumina samples with porosities between 10 and 44% and the mechanical properties were studied against the porosity [25]. Polymers (such as polymethyl methacrylate, polystyrene), low molecular organic materials (e.g. sucrose) and graphite have also been used as pore formers in preparing yttria stabilized zirconia ceramic tape cast products with well-defined porosities [26].

The third example is an attempt to obtain oxide dispersed steel (ODS), which contains finely dispersed oxide ceramic particles and is mostly prepared by mechanical alloying [27], followed by e.g. cold or hot isostatic pressing [28] or spark plasma sintering [29]. A group prepared ODS samples by the sol-gel route followed by high temperature reduction in argon-hydrogen gas mixture, the consolidation of the powder obtained was achieved by spark plasma sintering [30]. A review devoted to alternative preparation technologies [31] mentions further methods, however, the method suggested and tried by us has not yet been mentioned. It consists in adding various organic metal salts or metal-organic compounds to metal injection molding feedstock and to sinter the shaped body in reductive atmosphere after de-binding, where it is expected that the additives are transformed into finely divided metal oxide particles. It is to be noted that the use of organometallic additives (siloxanes and titanates) in CIM of SiN based systems has been described [32], but that is a completely different application, as there these additives were mainly used as surface treatment (coupling) agents [33, 34]. One article [35] mentions oxidative treatment of MIM processed steel, but there iron oxide is formed and not a new ceramic oxide is added to the system. In spite of the incomplete success of the method proposed by us (in terms of oxide size distribution) we are sure that further refinement of this approach will lead to success.

## 2. Experimental
### 2.1. Materials
#### 2.1.1. Heat resistant chimney lining prepreg

The detailed formulation of the prepreg sample studied will not be given for proprietary reasons, but it is closely related to the systems described in the examples of patent EP2826811 (A2) quoted above. The samples were provided by Polinvent Ltd. (Hungary).

### 2.1.2. Feedstock components and additives for preparing porous alumina by CIM

For the fabrication of the binder component of the CIM feedstock the following materials were used: Tipelin BS 501-17 high density polyethylene (HDPE) with hexane co-monomer, MFR = 0.2 g/10 min (TVK, Hungary), Plexar PX5335, an anhydride modified linear low density polyethylene (LLDPE) tie-layer resin, MFR = 5.7 g/10 min (Lyondell Basell, USA), GUR 4022-6 an UHMWPE (ultrahigh molecular weight polyethylene grade with a molecular mass of about 5 million) with coarse grain size, suggested for porous products by sintering (Celanese, USA), Luwax OP wax, melting point about 100 °C (BASF, Germany), oleic acid, 99% purity (Sigma Aldrich, Germany). From these the first two, the wax and the oleic acid components were the "regular" components of the binder, UHMWPE was explored as a potential pore forming agent (as it does not melt). In addition to UHMWPE we also tried citric acid 99% purity (Sigma Aldrich, Germany) as a pore forming agent, as it emits $CO_2$ during thermal decomposition. The inorganic components were alumina ($Al_2O_3$, ALO-G4-4G, MAL, Hungary) with $d_{50}$ = 4–7 μm particle size, specific surface 1.3 $m^2 \cdot g^{-1}$, and alumina trihydrate ($Al(OH)_3$, ALOLT-01, MAL, Hungary, $d_{50}$ = 40–80 μm) – the latter was tried as a pore forming agent, as it emits water vapor during decomposition.

### 2.1.3. Feedstock components and additives for preparing ODS steel by MIM

The stainless steel (SS) powder used in our experiments was kindly provided by Högenäs, Sweden (316 LHC, 17% Cr, 12% Ni, 2.2% Mo, Si content between 0.5 and 1.0%). According to a data leaflet the density of the SS316 alloy [36] is about 8.00 g/cm$^3$ (although it is interesting to note that the calculated density, assuming volume additivity and using the densities of the constituent element gives only 7.70 g/cm$^3$), while according to the data leaflet attached to the product, the density of a 600 MPa compacted powder with a wax lubricant is about 6.50 g/cm$^3$, which means about 18.8% porosity.

For the preparation of MIM feedstock binder material the same polymers (viz. Tipelin BS 501-17 and Plexar PX5335) and oleic acid were used, as in the previous example, but here the wax component was 005 E/AA oxidized PE wax (kindly provided by Honeywell, USA). The following additives were tried to produce ceramic particles after the debinding and sintering process: titanium (IV) isopropoxide (97%, Sigma Aldrich, Germany), tetraethoxysilane (98% Sigma Aldrich, Germany), aluminum mono-stearate (~75%, technical, Sigma Aldrich, Germany), KR® TTS (Titanium IV 2-propanolato, tris isooctadecanoato-O, Kenrich Petrochemicals Inc., USA), KA 322 (Di-isopropyl(oleyl)aceto acetyl aluminate, Kenrich Petrochemicals Inc., USA). Samples of Kenrich products were kindly provided by Farrl GmbH, Germany.

## 2.2. Sample preparation
### 2.2.1. Preparation of the cured and heat treated chimney lining prepreg samples

We received the impregnated and steam cured tube samples for testing from Polinvent Ltd. General methods for preparing such chimney linings are amply demonstrated in the examples of patent EP2826811 (A2). The inorganic filler samples are thoroughly dried, the organic binder components are mixed first at room temperature then the mixture is poured into a dissolver and homogenized at an elevated temperature (about 60 °C) and the fillers are gradually added with increasing rotation speed. The mixture is incubated at about 50 °C for a longer time (about 8 hours) under dry nitrogen blanket in the dissolver. A glass cloth hose (areal weight around 2000 g/m$^2$) is impregnated by the resin mixture, which is still flowable at this time. A polyethylene tubular film is placed into the resin impregnated reinforcement hose. The structure is inflated from inside the polyethylene tube by about 100 °C steam for about 1–1.5 hour.

In addition to the steam cured samples we have studied further samples: post-cured at 195, 200 and 205 °C for 60 minutes (to explore the effect of post-curing temperature); heat treated at 300 °C for 60 minutes. Parts cut from the samples post-cured at 200 and 300 °C were exposed to further, extreme heat treatment: 5 °C/min heating from room temperature to 300 °C followed by 10 min isotherm; 1 °C/min heating to 400 °C followed by 10 min isotherm; 5 °C/min heating to 600 °C, spontaneous cooling. This pyrolytic treatment was made in air atmosphere, as the flue gases also contain oxygen. (This upper temperature was selected because, according to the thermal analytical data to be described hereafter there was no essential change in the weight and no thermal effects were observed beyond this temperature).

### 2.2.2. Preparation of the CIM and MIM feedstock samples

As the powder injection feedstock samples in the second and third examples were prepared by very similar methods, therefore they will be described together. The feedstock components were mixed in a Brabender Plasti-Corder (Brabender GmbH, Germany) equipped with a kneading chamber having two Z-type mixing elements, capable of mixing about 150 cm$^3$ material, and oil heating (up to 220 °C). Preparation of some batches selected for injection experiments was repeated with larger kneading chamber as well. After adding a minimal amount of mineral or the metal powder to the pre-heated chamber about 80% of the polymer component was added within about 5 min to the chamber at 5 rpm (rotation per minute) kneading speed: first the PE polymer, then the copolymer component, followed by the addition of about 75% of the solid powder (ceramic or metal) in about 10 minutes, maintaining the 5 rpm speed, then the other low molecular components were added, followed by the rest of the solid powder (5 min) and the rest of the polymer components (20%). The total feed time was 20 minutes, then the rotation speed was increased to 20 rpm and the mixing was continued for 10 minutes. The still hot mixture was removed from the mixing chamber and placed on an aluminum tray. The composition of the CIM feedstock samples prepared is described in Table 1. The volume percent of the inorganic components was around 70% in compounds C01-C08, while it varied between 50 and 65% in the rest of the compounds. With the exception of samples C08-C12 which were homogeneous or contained large chunks of continuous paste, others were more powder-like just after compounding. (Figure 1 shows as an example a photograph of the freshly prepared feedstock samples of C01 and C09 exhibiting powder-like/granular and continuous structures respectively). We tried to explore the effects of homo/copolymer ratio, polymer/wax ratio and the content of the potentially pore forming compound (citric acid, ATH and UHMWPE) on compoundability and on final morphology. The still hot feedstock was processed further by compression molding (in the case of the MIM samples we used only this method) using a hydraulic Bucher Guyer KHA press. First a preliminary compression force was applied at 220 °C at low pressure (less than 1 ton). This resulted in a press-form of about 1 mm thickness, which was processed further in a square mold (160×160×3 mm) using the following compression program: 0 tons for 10 minutes, 2 tons for 5 minutes, 20 tons for 3 minutes. (Figure 2 shows the photograph of compression molded CIM feedstock samples C01 and C11, exhibiting granular and continuous structures before compression molding respectively).

**Table 1.** Composition of the alumina based ceramic injection molding (CIM) feedstock samples in weight% (the actual grades are identified in section 2.1.2).

| [wt%] | C01 | C02 | C03 | C04 | C05 | C06 | C07 | C08 | C09 | C10 | C11 | C12 | C13 | C14 |
|---|---|---|---|---|---|---|---|---|---|---|---|---|---|---|
| HDPE | 3.0 | 3.0 | 1.2 | 3.0 | 3.0 | 3.0 | 3.0 | 2.9 | 4.8 | 9.5 | 4.0 | 3.4 | 2.8 | 3.1 |
| PE-copolymer | 3.0 | 3.0 | 3.0 | 3.0 | 3.0 | – | – | 2.9 | 4.6 | 9.8 | 3.9 | 3.3 | 2.7 | 2.6 |
| UHMW PE | – | – | 1.8 | – | – | 3.0 | – | – | – | – | – | – | – | – |
| OP Wax | 3.0 | – | 3.0 | 3.0 | 3.0 | – | 3.0 | 5.0 | 9.8 | – | 8.3 | 6.9 | 5.7 | 6.3 |
| Oleic acid | 0.6 | 0.6 | 0.6 | 0.6 | 0.6 | 0.6 | 0.6 | 0.6 | 0.7 | 0.7 | 0.7 | 0.7 | 0.6 | 0.6 |
| Al$_2$O$_3$ | 90.4 | 90.4 | 90.4 | 60.2 | 30.1 | 90.4 | 90.4 | 88.5 | 80.1 | 80.0 | 83.1 | 85.8 | 88.2 | 87.3 |
| ATH | – | – | – | 30.1 | 60.2 | – | – | – | – | – | – | – | – | – |
| Citric acid | – | 3.0 | – | – | – | 3.0 | 3.0 | – | – | – | – | – | – | – |

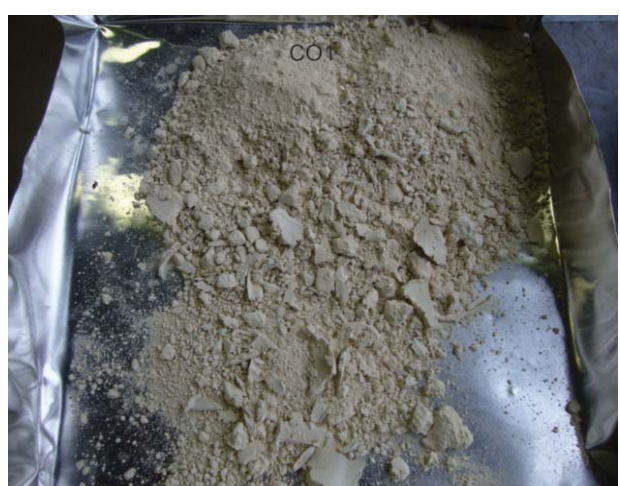 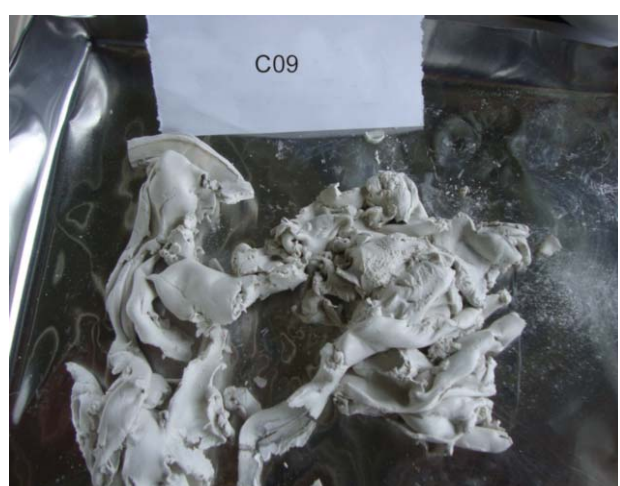

**Figure 1.** Photographs of the freshly prepared feedstock samples of C01 (granular structure, (a)) and C09 (continuous structure, (b)).

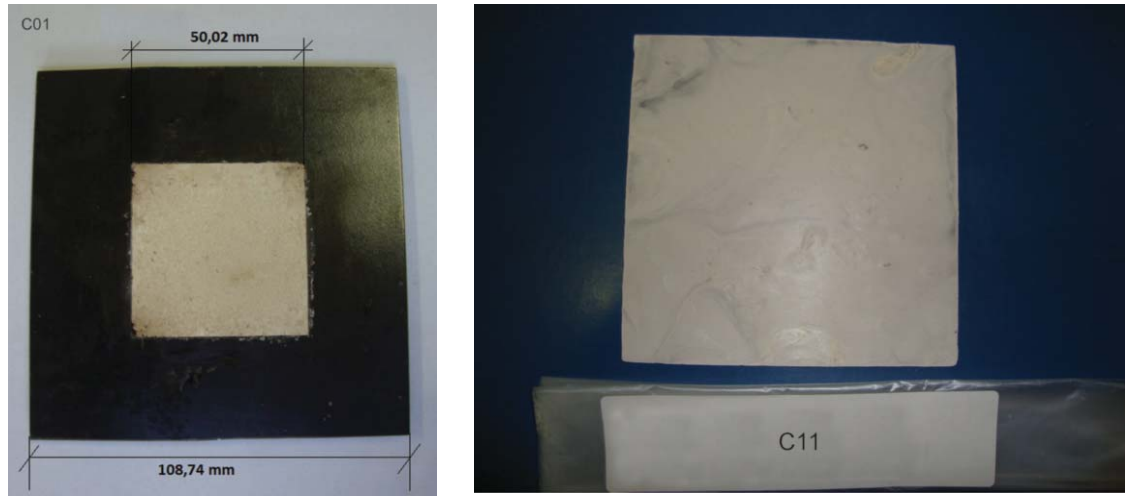

**Figure 2.** Photographs of compression molded feedstock samples C01 (granular structure, (a)) and C11 (continuous structure, (b)). Figure 2(a) shows the metal frame too used for compression molding.

Some selected CIM feedstock materials (namely C09 and C11 of Table 1, produced in larger quantities) were processed further with laboratory size plunger-torpedo type injection molding machine [37] located on a laboratory scale combined processing unit (Machine Combinée de Laboratoire / Presse injection, Cie AMIL, France) used at the Miskolc University for education purposes, as this type of equipment is less sensitive to the homogeneity of the mixture than standard screw based injection units. This injection unit is vertically oriented and can be operated manually with high pressure air. The temperature of the injection unit was set to 220 °C. For this purpose the compression molded CIM feedstock was cut into smaller pieces using a heavy duty Piovan granulator equipped with rotating knives (RSP1515, Piovan, Italy).

**Table 2.** Composition of the stainless steel powder based metal injection molding (MIM) feedstock samples in weight% (the actual grades are identified in section 2.1.2). The last row shows the solvent used for dispersing the metal-organic compound on the stainless steel (SS) powder before compounding. iPrOH = isopropanol, THF = tetrahydrofurane.

| [wt%] | M01 | M02 | M03 | M04 | M05 | M06 | M07 |
|---|---|---|---|---|---|---|---|
| HDPE | 1.1 | 1.7 | 1.1 | 1.1 | 1.1 | 1.5 | 1.1 |
| PE-copol | 1.1 | – | 1.1 | 1.1 | – | – | – |
| Ox PE Wax | 1.1 | 1.7 | 1.1 | 1.1 | – | – | – |
| Oleic acid | 0.2 | – | 0.2 | 0.2 | – | – | – |
| SS powder | 96.6 | 96.6 | 95.9 | 95.9 | 95.6 | 95.1 | 95.8 |
| Ti(iPrO)$_4$ | – | – | 0.7 | – | – | – | – |
| TEOS | – | – | – | 0.7 | – | – | – |
| Al-stearate | – | – | – | – | 3.3 | – | – |
| KR® TTS | – | – | – | – | – | 3.5 | – |
| KA322 | – | – | – | – | – | – | 3.1 |
| Solvent | iPrOH | iPrOH | iPrOH | iPrOH | iPrOH | THF | iPrOH |

The composition of the prepared MIM feedstock samples is given in Table 2. Here the sample preparation process was somewhat different from the CIM feedstocks. Samples M01 and M02 were used to check whether all three organic components (HDPE, PE-copolymer and the oxidized PE wax) are needed to prepare the MIM feedstock or two are enough. Samples M03-M07 contained various metal-organic additives which were designed to give similar amounts (about 0.3 wt%) of alumina, silica or titania particles provided that their Al, Si and Ti content is transformed to the respective oxides during the de-binding and sintering process. In the case of three additives: Al-stearate, KR® TTS and KA322 which contain long chain aliphatic moieties, thus it can be expected that they play a lubricating role during feedstock compounding, no PE-copolymer or oxidized PE wax component was used. Samples M01-M07 contained about 70–77 vol% stainless steel powder.

The compounding process was similar to that of CIM feedstock samples described above, the only difference was that the additives shown in Table 2 for compounds M03–M07 were previously mixed with the SS powder in a solvent. The calculated amount of the additive was mixed with 100 ml solvent in a 250 ml round bottom flask for 1 hour at room temperature, then 450 g SS powder was added and the mixing process was continued for another hour. Then the solvent was distilled from the mixture and the residue was dried in a vacuum oven until constant weight at 80 °C. If the "coated" SS powder became clumped, it was first broken down by a pestle in a mortar before adding it to the already molten polymer components (as described for the CIM feedstock compounding process). It has to be noted, however, that at the temperature of compounding eventual aggregates broke down easily. The freshly prepared MIM feedstock has been compression molded to 50×50×5 mm samples using a pressing frame, at 220 °C, 20 tons. Other details of the compression molding process were identical with that described for CIM feedstocks. The compression molded plates were cut into about 10 mm wide stripes to fit into the sample holder of the de-binding and sintering oven with controlled atmosphere. (In this case we did not try injection molding, as we were interested in the morphology only).

### 2.2.3. De-binding and sintering process

The removal of the organic components (de-binding) from the "green body" to obtain the "brown body" was done in air for the CIM compounds in a custom made oven in ambient atmosphere. The following temperature program was used: 2 °C/min heating from room temperature to 280 °C followed by 30 min isotherm, then 0.5 °C/min heating to 400 °C followed by 60 min isotherm, 1 °C/min heating to 550 °C followed by 120 min isotherm. Cooling was spontaneous with a time constant of about 250 minutes.

The sintering of the CIM brown bodies (both the compression molded and the injection molded ones) was also performed in air using the following temperature program: heating by 5 °C/min form room temperature to 900 °C followed by 30 min isotherm, 5 °C/min heating to 1600 °C followed by 120 min isotherm. Cooling was spontaneous with a time constant of about 250 minutes.

The de-binding and sintering processes of the (previously compression molded) MIM green bodies were performed in another custom-built oven that can be evacuated or flushed with inert gas, in argon gas atmosphere with the addition of 3 vol% hydrogen to prevent the oxidation of steel particles. (It was assumed that the oxygen content of the metal-organic additives is enough to transform the additives to the corresponding oxides during the initial phase of de-binding). At first the oven was heated spontaneously to 80 °C followed by 8 min isotherm, and evacuation. The 97% Ar – 3% $H_2$ gas flush was switched on and the spontaneous heating to 120 °C was again followed by 8 min isotherm, and evacuation. This was repeated again by spontaneous heating to 160 °C followed by 8 min isotherm, and evacuation. This sequence was used to remove all air and water from the samples. Thereafter the reductive/inert gas flush was maintained until the sample cooled back to room temperature at the end. Under this constant gas flow the following heating sequence was applied: 0.5 °C/min heating to 400 °C followed by 30 min isotherm; 1 °C/min heating to 550 °C followed by 60 min isotherm; 1 °C/min heating to the sintering temperature followed by 60 min isotherm; spontaneous cooling. The sintering temperature for the compounds M01 and M02 was varied between 950 and 1300 °C and from density tests to be described later the optimum sintering temperature was decided to be 1150 °C.

### 2.3. Test methods

For the TGA/DTA tests a Setaram Setsys 16/18 (France) equipment was used, using a heating rate of 10 °C/min from 30 to 1000 °C in argon atmosphere, flush rate was 30 ml/min. The densities were measured by a RADWAG-PS210-R2 type Archimedean balance (Poland) using water as buoyancy fluid. A part of the SEM micrographs was taken with a Hitachi TM-1000 equipment using Solid State Backscattered Electron Detector (BSE), some others (together with EDX) were taken on a Jeol JSM-6380LA Analytical Scanning Electron Microscope (Japan). Optical micrographs were taken on a Keyence VHX-2000 type digital microscope. X-ray diffractograms were measured on Rigaku MiniFlex II desktop equipment. X-ray photoelectron spectra were recorded on a Kratos XSAM 800 spectrometer operated at fixed analyzer transmission mode, using Mg K$\alpha_{1,2}$ (1253.6 eV) excitation. The high resolution photoelectron lines of the main constituent elements were recorded by 0.1 eV steps. Spectra were acquired and processed by the Kratos Vision 2 software package. Area intensity data were obtained after Shirley type background removal. Quantitative analysis was performed by the XPS MultiQuant 7.7 program [38] using the experimentally determined photo-ionisation cross-section data of Evans *et al.* [39] and asymmetry parameters of Reilman *et al.* [40]. Raman Spectra were measured on a Labram type Raman microscope (Horiba-Jobin Yvon), to which a diode laser manufactured by Sacher Lasertechnik was attached (785 nm, ~100 mW nominal power). Spectra were recorded in the 200–3000 cm$^{-1}$ range. The test time was: 2×40 sec. The grating used had 950 grooves/mm. The magnification of The objective was 10× with a confocal hole of 1000 μm. The spectral resolution was ~5 cm$^{-1}$. Transmission FTIR spectra in KBr pellet were measured on a Bruker Tensor 27 spectrometer using 16 scans 2 cm$^{-1}$ resolution.

## 3. Results and discussion
### 3.1. Heat resistant chimney lining prepreg

The patent EP2826811 (A2) describes the main features of the assumed chemical processes occurring during the curing, post-curing processes and under application conditions, at high temperatures. The prepreg tube to be pulled into the chimney is protected from inside by a polyethylene tubular film. The primary pre-curing of the prepreg system is initiated by 100 °C steam from inside, which permeates through the polyethylene tubular film and starts to react with the isocyanate component, yielding carbamic acid and finally aromatic amine and carbon dioxide. The $CO_2$ gas formed during this reaction reacts with the fine metal hydroxide powder resulting in metal carbonates and hydrocarbonates. The freshly formed water drives the curing reaction further. In the isocyanate – water reaction not only $CO_2$ is released, but also aromatic amines are formed, which react with the residual isocyanate resulting in polyurea formation. As the prepreg is heated up it softens and, until the crosslinking prevents flow, it impregnates thoroughly the reinforcing fiber fabric hose under the effect of the mechanical pressure of the hot steam. After the removal of the PE tubular film the reaction is completed at a curing temperature around 200 °C. The consolidated structure still contains ATH which, if exposed to a temperature around 300 °C, loses water and transforms partly or completely into $Al_2O_3$. At even higher temperatures the resin component itself undergoes degradation but, due to the char formation the whole structure remains self-supporting and can withstand substantially higher temperatures.

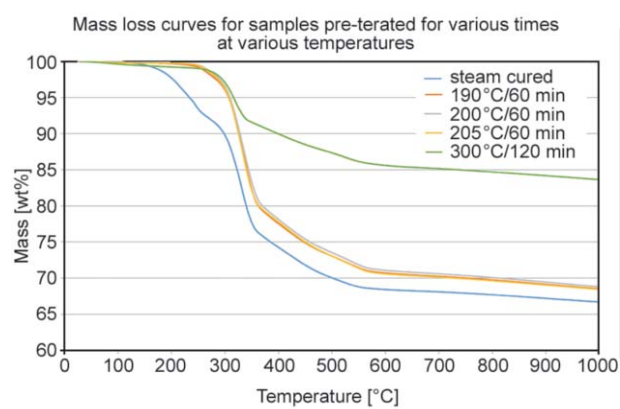

**Figure 3.** TGA curve of the prepreg sample K/80/WX3/43 exposed to various heat treatments.

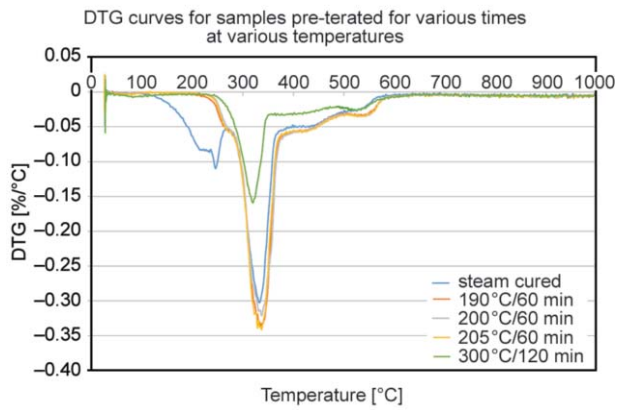

**Figure 4.** DTG curve of the prepreg sample K/80/WX3/43 exposed to various heat treatments.

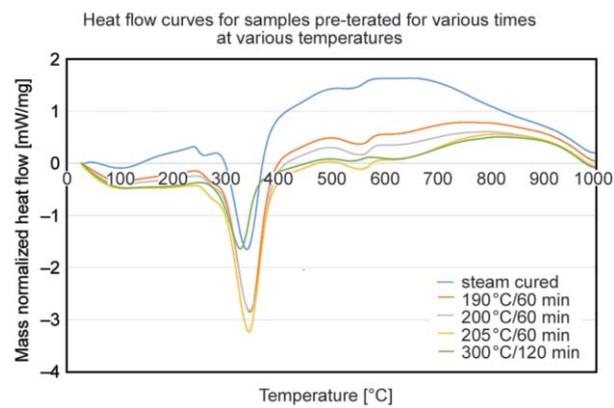

**Figure 5.** DTA curve of the prepreg sample K/80/WX3/43 exposed to various heat treatments.

First the transformation process was studied by simultaneous differential thermoanalysis (DTA) – thermogravimetry (TGA) on samples cured at 100 °C for 30 min using steam (steam cured), on samples post-cured for 30 min at 195, 200 and 205 °C to see whether these minor temperature changes cause any significant difference and on a sample heated to 300 °C for 120 min – to simulate a stage when the transformation of the ATH is almost complete. The %weight change (TGA), its derivative (DTG) and the temperature difference curve between the sample and an inert reference sample holder (DTA) are shown in Figures 3, 4 and 5 respectively. All three figures clearly show that the exact curing tempera-

**Table 3.** Characteristic transition temperatures observed on the DTG and DTA curves of steam pre-cured (100 °C, 60 min), cured (195, 200 and 205 °C, 30 min) and heat treated (300 °C, 120 min) prepreg samples shown in Figures 3–5. min. = minimum, sh = shoulder, sh? = uncertain shoulder

| Sample | $T_{(DTG)}$ [°C] | $T_{(DTA)}$ [°C] |
|---|---|---|
| Steam cured | – | 100 (min.) |
|  | 245 (min.) | 257 (min.) |
|  | 334 (min.) | 340 (min.) |
|  | 440 (sh) | 440 (sh) |
|  | 532 (sh) | 536 (sh) |
| 200 °C 60 min. | – | 99 (min.) |
|  | 275 (sh) | 277 (sh) |
|  | 335 (min.) | 345 (min.) |
|  | 442 (sh) | 429 (sh) |
|  | 540 (min.) | 550 (min.) |
| 300 °C 120 min. | 86 (min.?) | 100 (min.) |
|  | – | 187 (min.) |
|  | 320 (min.) | 322 (min.) |
|  | 403 (min.) | 412 (sh?) |
|  | 524 (min.) | 534 (min.) |
|  | – | 620 (min.) |

ture between 195 and 205 °C does not influence too much the process (in spite of the 10 °C temperature difference, which is significant in most chemical reactions), but there are important differences between the steam cured (100 °C), post-cured cured (200 °C) and heat treated (300 °C) samples. The difference between the 200 °C post-cured and the 100 °C steam cured samples reaches a maximum of about 6.5 wt% around 300 °C, but later the difference is reduced to about 2 wt%, which means that relatively few material loss occurs between the 100 °C curing and 200 °C post-curing processes. The weight loss difference between the 200 °C post cured and 300 °C heat treated samples reaches about 14.8 wt% in the high temperature range, which can be presumably ascribed to the decomposition of the ATH component. It is interesting to compare the DTA and the DTG curves up to 600 °C and higher (see Figures 4 and 5 and Table 3), as they show which are the decomposition processes accompanied by weight loss. The DTG minimum around 320 °C is due to the water loss of ATH (pure ATH exhibits a similar minimum under identical conditions at 320 °C – somewhat depending on the crystal structure). The steam cured sample exhibits a loss minimum around

*Bánhegyi et al.: Monitoring Thermal Transformations in Hybrid Composites***Table 4.** Surface concentration (atomic %) of various elements determined by X-ray photoelectron spectroscopy on steam pre-cured (100 °C, 60 min), cured (200 °C, 30 min) and heat treated (300 °C, 120 min) prepreg samples and the last two after gradual heating to 600 °C (heating program is given in the Experimental part). Carbon is decomposed into various chemical states.

| Atomic% | O | N | C–H C–C | C–O C–N | O–C–O | Al | S | P | Si | Na |
|---|---|---|---|---|---|---|---|---|---|---|
| 100 °C | 17.7 | 9.5 | 49.7 | 14.5 | 6.4 | 1.0 | 0.6 | 0.2 | 0.4 | – |
| 200 °C | 18.3 | 9.0 | 52.9 | 10.9 | 5.7 | 1.4 | – | 1.0 | 0.0 | 0.8 |
| 300 °C | 9.6 | 3.5 | 75.6 | 7.8 | 2.1 | 1.1 | 0.3 | 0.1 | 0.0 | – |
| 200–600 °C | 57.4 | 1.7 | 3.4 | 4.9 | 0.9 | 24.2 | – | 3.6 | 1.5 | 2.3 |
| 300–600 °C | 50.2 | 2.1 | 8.5 | 10.6 | 1.6 | 24.8 | – | 0.0 | 1.2 | 1.0 |

245 °C, which is presumably due to the incomplete curing of the isocyanate, which is completed during the post-curing process around 200 °C. The sample heat treated at 300 °C still contains some residual ATH, as shown by the first weight loss minimum. The higher temperature, less intense weight losses accompanied by weak endotherms are presumably due to the thermal decomposition of the organic components. It is interesting to observe that (especially in the steam cured sample) there is a broad exothermal background on the whole DTA curve which gradually diminishes in the cured and post-cured samples (although it can be due to the background correction process).

We tried to understand – at least partially – the chemical transformation process using two methods that can measure the changes in the elemental composition of the samples in various stages of the process. From these XPS (X-ray photoelectron spectroscopy) probes the uppermost few nanometers of the sample, while the penetration depth of SEM/EDX depends somewhat on the acceleration voltage, but it is typically in the order a few micrometers (see e.g. [41]). Table 4 summarizes the atomic% of the surface composition of the steam cured, the 200 °C post-cured and 300 °C heat treated samples and the last two samples after gradually heating them to 600 °C (see the Experimental part for the heating program). In the case of the carbon the C1s peak was resolved into components belonging to C–H, C–C, C–O/C–N and O–C–O bonded carbon atoms. Relevant chemical shift values can be found in Ref. [42] dealing with the XPS and ToF SIMS investigations of aluminum surface and polymeric MDI. Due to the partial oxidation and hydration of the Al surface almost all species relevant to our study are present. Some features can be well recognized. The differences between the steam cured and the 200 °C cured samples are minimal. The surface of the steam cured sample is clearly dominated by the organic component. The N/C atomic ratio is not far from that of MDI (the main organic component), the O/C ratio is somewhat larger, presence of P and S is due to minor additives. The aluminum component of ATH hardly appears on the surface. The 200 °C post-curing slightly increases the O/C ratio and reduces the N/C ratio, otherwise there are minimal differences. The 300 °C heat treatment (wherein the major part of ATH is decomposed, as shown by the TGA/DTA results shown above) results in a drastic reduction of the O/C and N/C ratios, which is also reflected by the fact that the relative weight of the carbon components attached to high electronegativity elements is significantly reduced. Interestingly the Al component still hardly appears on the surface. Presumably the polyurea network begins to degrade at this temperature and new, polyaromatic structures are formed.

**Table 5.** Atomic% composition of the steam-cured (100 °C), cured (200 °C), heat treated (300 °C) and the calcined (200 to 600 °C and 300 to 600 °C) prepreg samples measured by SEM/EDX at various points of the samples.

| Atomic% | Spot | C | N | O | Al | P | Si | Na | Ca |
|---|---|---|---|---|---|---|---|---|---|
| 100 °C | 1 | 57.3 | 9.9 | 26.9 | 4.7 | 1.3 | – | – | – |
|  | 2 | 42.7 | 5.9 | 42.2 | 8.7 | 0.5 | – | – | – |
|  | 3 | 2.9 | 1.7 | 65.2 | 30.3 | – | – | – | – |
|  | 4 | 42.2 | 4.7 | 43.2 | 9.0 | 0.8 | – | – | 0.1 |
| 200 °C | 1 | 14.3 | 1.5 | 56.6 | 27.7 | – | – | – | – |
|  | 2 | 44.5 | 10.3 | 38.0 | 7.2 | – | – | – | – |
|  | 3 | 43.7 | 7.8 | 39.6 | 8.9 | – | – | – | – |
|  | 4 | 8.9 | – | 68.6 | 22.6 | – | – | – | – |
| 300 °C | 1 | 4.6 | – | 66.0 | 29.4 | – | – | – | – |
|  | 2 | 44.2 | 7.7 | 37.0 | 8.8 | – | 2.2 | 0.1 | – |
|  | 3 | 5.5 | 1.7 | 66.8 | 25.9 | – | 0.1 | 0.2 | – |
|  | 4 | 34.7 | 4.6 | 45.4 | 14.8 | 0.4 | – | 0.1 | – |
| 200 to 600 °C | 1 | 7.9 | – | 58.3 | 33.7 | 0.1 | – | – | – |
|  | 2 | 9.3 | – | 56.0 | 34.7 | 0.1 | – | – | – |
|  | 2 | 9.0 | – | 40.0 | 51.0 | – | – | – | – |
|  | 4 | 12.2 | – | 57.4 | 30.4 | – | – | – | – |
| 300 to 600 °C | 1 | 5.4 | – | 49.9 | 44.7 | – | – | – | – |
|  | 2 | 11.4 | – | 58.5 | 30.1 | – | – | – | – |

**Table 6.** Atomic% ratios determined for the steam-cured (100 °C), cured (200 °C), heat treated (300 °C) and the calcined (200 to 600 °C and 300 to 600 °C) prepreg samples measured by XPS and SEM/EDX methods. In the latter case carbon-rich and aluminum rich areas are distinguished wherever appropriate and only ranges are given due to the local variations.

|   |   | O/C (XPS) | O/C (EDX) | N/C (XPS) | N/C (EDX) | C/Al (XPS) | C/Al (EDX) | O/Al (XPS) | O/Al (EDX) |
|---|---|---|---|---|---|---|---|---|---|
| 100 °C | C rich | 0.3 | 0.5–1.0 | 0.1 | 0.1–0.2 | 70.6 | 5–12 | 17.7 | 5–6 |
|  | Al rich |  | 23 |  | 0.6 |  | 0.1 |  | 2.1 |
| 200 °C | C rich | 0.3 | 0.8–0.9 | 0.1 | 0.2 | 49.6 | 5–6 | 13.1 | 4–5 |
|  | Al rich |  | 4–8 |  | 0–0.1 |  | 0.4–0.5 |  | 2–3 |
| 300 °C | C rich | 0.1 | 0.8–1.3 | 0.0 | 0.15 | 77.7 | 2–5 | 8.7 | 3–4 |
|  | Al rich |  | 12–14 |  | 0–0.3 |  | 0.2 |  | 2–3 |
| 200–600 °C | Al rich | 6.2 | 4–8 | 0.2 | 0 | 0.4 | 0.2–0.4 | 2.4 | 1–2 |
| 300–600 °C | Al rich | 2.4 | 5–9 | 0.1 | 0 | 0.8 | 0.1–0.4 | 2.0 | 1–2 |

The really important changes accompany the high temperature treatment to 600 °C. The carbon content is drastically reduced, the Al component becomes prominent (the O/Al ratio is close to 2, or even higher, which is higher than 1.5 expected for alumina, which shows that other oxygen containing components may be present).

The elemental composition has been measured by SEM/EDX as well at various spots on the sample surface after various heat treatments, the results are summarized in Table 5, while Table 6 compares some atomic% ratios determined by XPS and SEM. It has to be noted that XPS and SEM/EDX cannot be compared directly not only because of the different penetration depths but also because of the fact that EDX is not really sensitive to low atomic numbers. Nevertheless the tendencies can be well interpreted and in general they are in agreement with the XPS observations. In the XPS tests of the steam cured sample Al could be hardly seen, while in the EDX results we can see carbon-rich and aluminum-rich areas. In the aluminum-rich areas the O/Al ratio is between 2 and 3 (with the exception of the calcined samples heated to 600 °C), where it is reduced to the range between 1 and 2. In ATH the expected O/Al ratio is 3, in alumina it is 1.5. Values below this limit indicate lower oxidation states. Nitrogen cannot be detected by EDX in the calcined samples, although XPS can still detect it and the N/C ratio measured by XPS does not change significantly at the highest treatment temperatures. Due to the deeper penetration the O/C ratios determined by EDX are higher than the XPS values – as more aluminum-bound oxygens are seen by this method even in the carbon-rich areas. The same difference explains the higher C/Al ratios measured by XPS than by EDX.

In addition to the XPS and SEM/EDX measurements vibrational spectroscopic studies were also made. Reflection Raman measurement could be performed only on the steam cured (100 °C) sample (see Figure 6) – all other samples produced so strong fluorescence background that it was not possible to detect a meaningful spectrum. Transmission FTIR spectra were detected for the steam cured (100 °C), the 300 °C heat treated and for the calcined (heated to 600 °C after the 300 °C treatment) samples (see Figure 7(a)–(c) in various spectral ranges). It has to be noted that the Raman spectrum did not contain any significant band above 1610 cm$^{-1}$ thus it is not really useful for the interpretation – as the skeletal vibrations are notoriously complicated to assign. Fortunately, however, the FTIR data are rich and informative. The OH/CH stretching region is dominated by the OH multiplet of the ATH filler, (where the exact location and relative intensity of the sub-bands depend on the gradually developing mineral form [43]) in the steam cured sample, and to a lesser degree, but still significantly in the 300 °C heat treated sample, while only a broad associated OH band can be seen in the calcined (300 to 600 °C) sample. Weaker bands in the 3400–3200 cm$^{-1}$ range may be due to the NH band of the amides groups formed. Very weak aromatic CH bands can be observed between 3000 and 3100 cm$^{-1}$ and aliphatic CH stretching bands between 3000 and 2800 cm$^{-1}$. Some remnants of the CH bands can be observed even in the calcined sample. In the carbonyl stretching region two bands can be observed in the steam cured sample at 1783 and 1739 cm$^{-1}$. The latter can be urethane carbonyl vibration, but the frequency of the first one is unusually high. In an aromatic polyurethane-urea based on MDI and a functionalized polystyrene polyol chain extended with a diamine [44] the 1775 and 1732 cm$^{-1}$ bands were assigned to amorphous and crystalline urethane vibrations respectively, while the 1602 cm$^{-1}$ band was attributed to the urea group. It is possible, however, that 1783 cm$^{-1}$ band in our system is due to uretdione formation [45] (self-dimerization of MDI). Formation of isocyanurates is also possible, but it would appear at 1690 cm$^{-1}$. (It is possible that the 1666 cm$^{-1}$ band observed in the steam cured sample is due to isocyanurates – but it can also be an amide band). The bands appearing at 1598, 1537 and 1511 cm$^{-1}$ are aromatic stretching vibrations well known from the MDI spectrum [46]. In the sample heat treated at 300 °C the OH bands of ATH are reduced, the CH stretching bands are very much reduced, the carbonyl vibrations practically disappear – only the aromatic skeletal vibrations remain with lower intensity. In the sample treated at 300 °C and in the same sample calcined at 600 °C the broad band appearing around 1600 cm$^{-1}$ is most probably due to adsorbed water. At 600 °C practically only the bands attributable to the inorganic components cold be detected.

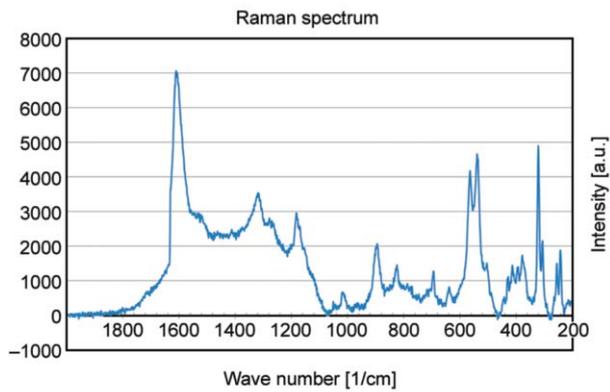

**Figure 6.** Raman spectrum of the steam cured (100 °C) prepreg sample K/80/WX3/43.

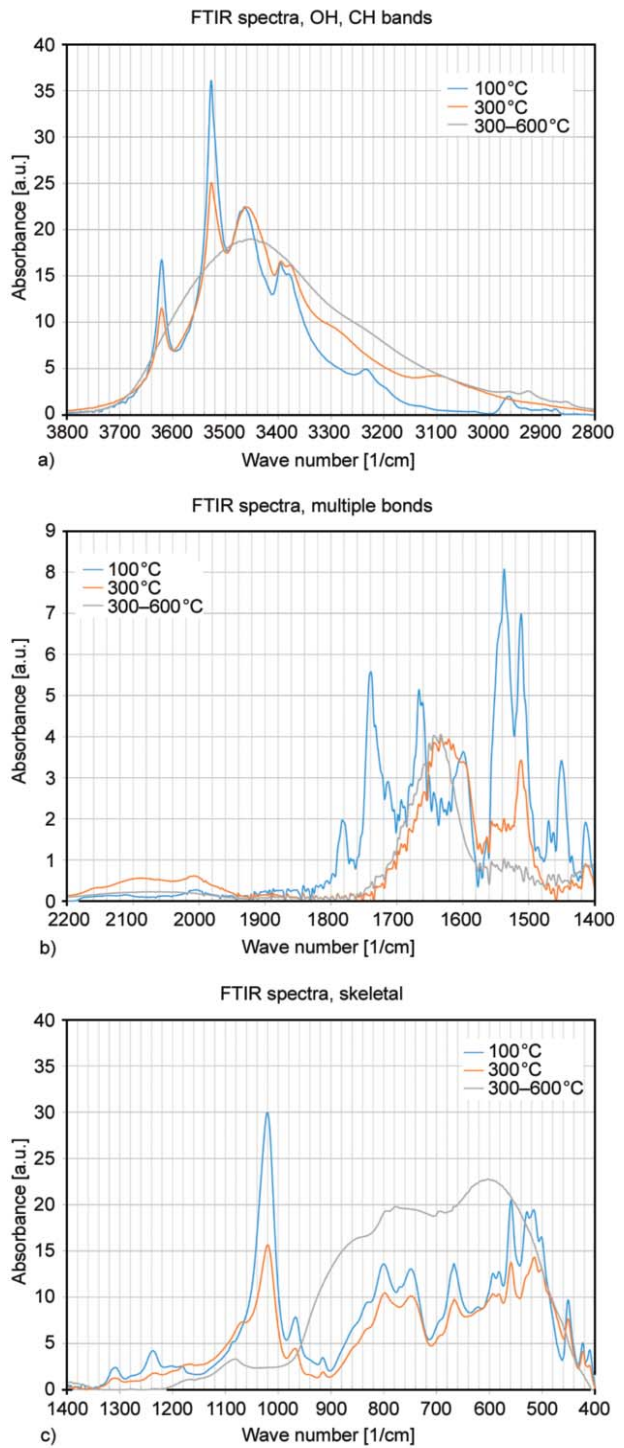

**Figure 7.** Transmission FTIR spectra of the steam cured (100 °C), 300 °C heat treated and 600 °C calcined (after 300 °C heat treatment) prepreg sample K/80/WX3/43 in the OH/CH stretching range (a), in the multiple bonds stretching range (b) and in the skeletal vibration range (c).

### 3.2. CIM compounds for preparing porous alumina samples

For samples C01-C04 and C07 we determined the experimental density of the "green body" by Archimedean principle (samples C05 and C06 disintegrated in water), after thermal de-binding the same was attempted, but mostly without success (most of the samples disintegrated in water). All samples could be sintered, however, successfully and the density was again determined for samples C01–C07. This allowed the calculation of the total porosity from the theoretical and experimentally determined densities, as shown by Equations (1) and (2):

$$\rho_{theoretical} = \frac{\sum_{i=1}^{n} m_i}{\sum_{i=1}^{n} \frac{m_i}{\rho_i}} \quad (1)$$

**Table 7.** The densities of the components used for calculating the theoretical densities and porosities of the CIM compounds presented in Table 1.

| Compound | Density [g/cm$^3$] |
|---|---|
| HDPE | 0.95 |
| PE-copol | 0.921 |
| UHMWPE | 1.00 |
| OP Wax | 0.98 |
| Oleic acid | 0.895 |
| Al$_2$O$_3$ | 3.95 |
| ATH | 2.42 |
| Citric acid | 1.66 |

$$porosity\% = 100\left(1 - \frac{\rho_{measured}}{\rho_{theoretical}}\right) \quad (2)$$

where $\rho_{theoretical}$ is the density of mixture calculated from the masses ($m_i$) and densities ($\rho_i$) of the components and $\rho_{measured}$ is the actually determined density. The component densities used for the calculation are shown in Table 7. The experimental densities and the calculated porosities for the green body and for the sintered final product are listed in Table 8. It can be seen that already the compression molded green body contains significant porosity, which is (with few exceptions) reduced after de-binding and sintering. It is to be noted that even the comparative sample (C01) not containing any pore forming agent exhibits about 9% porosity after sintering comparable with samples C02, C04 and C07, while samples C03, C05 and C06 exhibit larger porosities in the range of 14–15%. The de-binding program has been selected so that all organic components and the ATH components of the CIM compounds decompose at the highest temperature

**Table 8.** Experimentally determined densities and calculated porosities of CIM samples (presented in Table 1) in the "green body" stage, and after sintering. (n.a. = not applicable, C05 and C06 green bodies disintegrated in water).

| Compound | Green body density (measured) [g/cm$^3$] | Green body calculated porosity [%] | Sintered body density (measured) [g/cm$^3$] | Sintered body calculated porosity [%] |
|---|---|---|---|---|
| C01 | 2.46 | 18.7 | 3.59 | 9.1 |
| C02 | 2.92 | 7.1 | 3.56 | 9.9 |
| C03 | 2.57 | 15.3 | 3.41 | 13.7 |
| C04 | 2.12 | 19.7 | 3.68 | 6.8 |
| C05 | n.a. | n.a. | 3.32 | 15.9 |
| C06 | n.a. | n.a. | 3.34 | 15.4 |
| C07 | 2.73 | 13.7 | 3.61 | 8.6 |

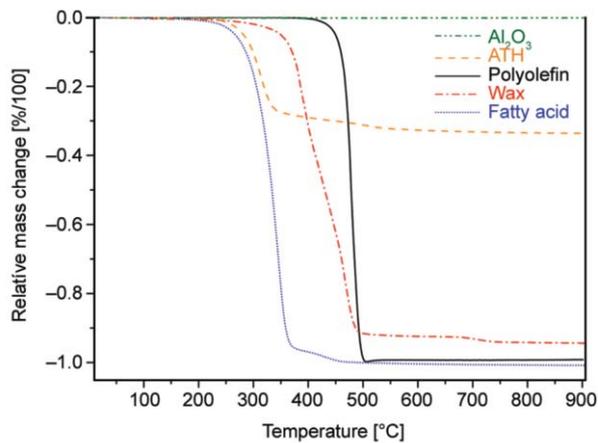

**Figure 8.** Typical TGA curves of the generic components used in preparing the CIM feedstock samples (see Table 1 for the components).

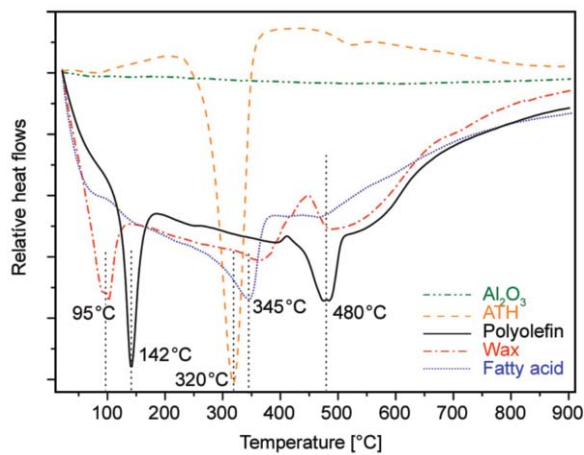

**Figure 9.** Typical DTA curves of the generic components used in preparing the CIM feedstock samples (see Table 1 for the components).

(500 °C). This conclusion came from the investigation of the TGA/DTA curves measured on separate components (see Figures 8 and 9). The DTA curves show the melting processes as well as the decomposition processes, while TGA only the latter. Therefore it is expected that the experimentally measured weight loss after de-binding should be equal to the weight% of the organic components (in the case of ATH containing compounds the water loss of the ATH components should also be taken into account). Theoretical and experimentally measured weight losses after de-binding are compared in Table 9. (In some cases, probably due to unavoidable chipping, the ratios of the measured and theoretical losses were higher than 100% – these are not indicated). It can be seen that in several cases (e.g. C04, C06, C10) the actual weight loss is only about 50% of the expected values are observed – indicating that there is some residual organic or otherwise volatile material which disappears only during the sintering step. Samples C01–C04 and C06–C07 preserve their integrity after sintering, while sample C05 (containing more ATH than $Al_2O_3$) exhibited fissures (probably too much volatile evolved). In samples C08–C14 we tried to reduce the $Al_2O_3$ content to improve the moldability (feedstock samples C08–C12 in fact exhibited larger, non-powdery chunks after mixing), but unfortunately the compounds which could be successfully injection molded by the injection molding machine available to us (C09 and C11) exhibited volume swelling and shape deformation during de-binding which remained after sintering (see Figure 10). Probably an injection molding machine specifically designed for CIM with much higher melt pressure could process directly the formulations C01–C07 too, which do not exhibit swelling and deformation on de-binding and sintering.

**Table 9.** Experimentally and theoretically determined weight losses in the CIM samples (presented in Table 1) after de-binding.

|     | Wt loss theor | Wt loss measured | Wt loss measured/Wt loss theor |
|-----|---------------|------------------|--------------------------------|
| C01 | 9.6%          | 8.0%             | 83.0%                          |
| C02 | 9.6%          | 6.1%             | 63.1%                          |
| C03 | 9.6%          | 8.1%             | 84.1%                          |
| C04 | 23.9%         | 12.9%            | 53.7%                          |
| C05 | 38.2%         | 23.3%            | 61.1%                          |
| C06 | 9.6%          | 4.3%             | 45.1%                          |
| C07 | 9.6%          | 6.8%             | 70.6%                          |
| C08 | 11.5%         | 8.4%             | 73.2%                          |
| C09 | 19.9%         | 25.0%            | –                              |
| C10 | 20.0%         | 9.8%             | 48.7%                          |
| C11 | 16.9%         | 17.9%            | –                              |
| C12 | 14.2%         | 17.0%            | –                              |
| C13 | 11.8%         | 11.0%            | –                              |
| C14 | 12.7%         | 14.0%            | –                              |

The morphology of the samples has been investigated by scanning electron microscopy. Figure 11 shows the surface of sintered samples of C01 and C04 (the former containing only alumina, the latter ATH too) as an example at various magnifications. The low resolution SEM micrographs show that the

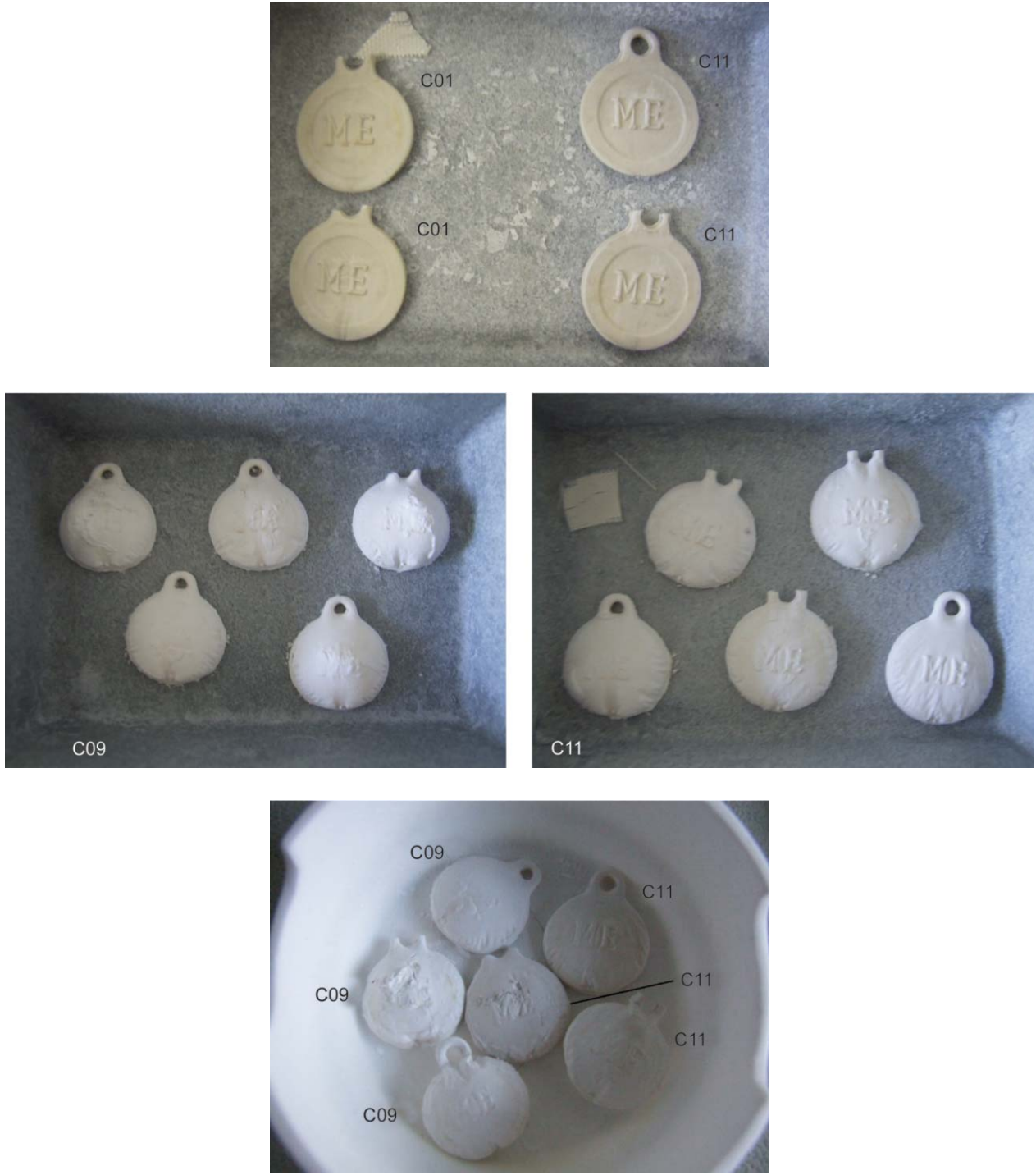

**Figure 10.** Photographs of injection molded feedstock samples C09 and C11 (a), C09 and C11 samples (b) after de-binding, and C09 and C11 samples after sintering (c). Samples flat in the "green" stage become swollen after de-binding, du to do not deform further after sintering.

surface of C04 is less regular, otherwise the two samples are fairly similar, they consist of particles of about 10 μm size an smaller sintered together at the contact points. Figure 12 shows SEM micrographs for samples C05 and C07 containing pore forming agents. Comparing C04 and C05 containing alumina and ATH in different ratios, it can be seen that an increased amount of ATH produces larger pores. The final porosity can be decreased by longer sintering time but if the goal is to produce porous bodies, it can be influenced by the amount of the pore forming agent and sintering conditions. An

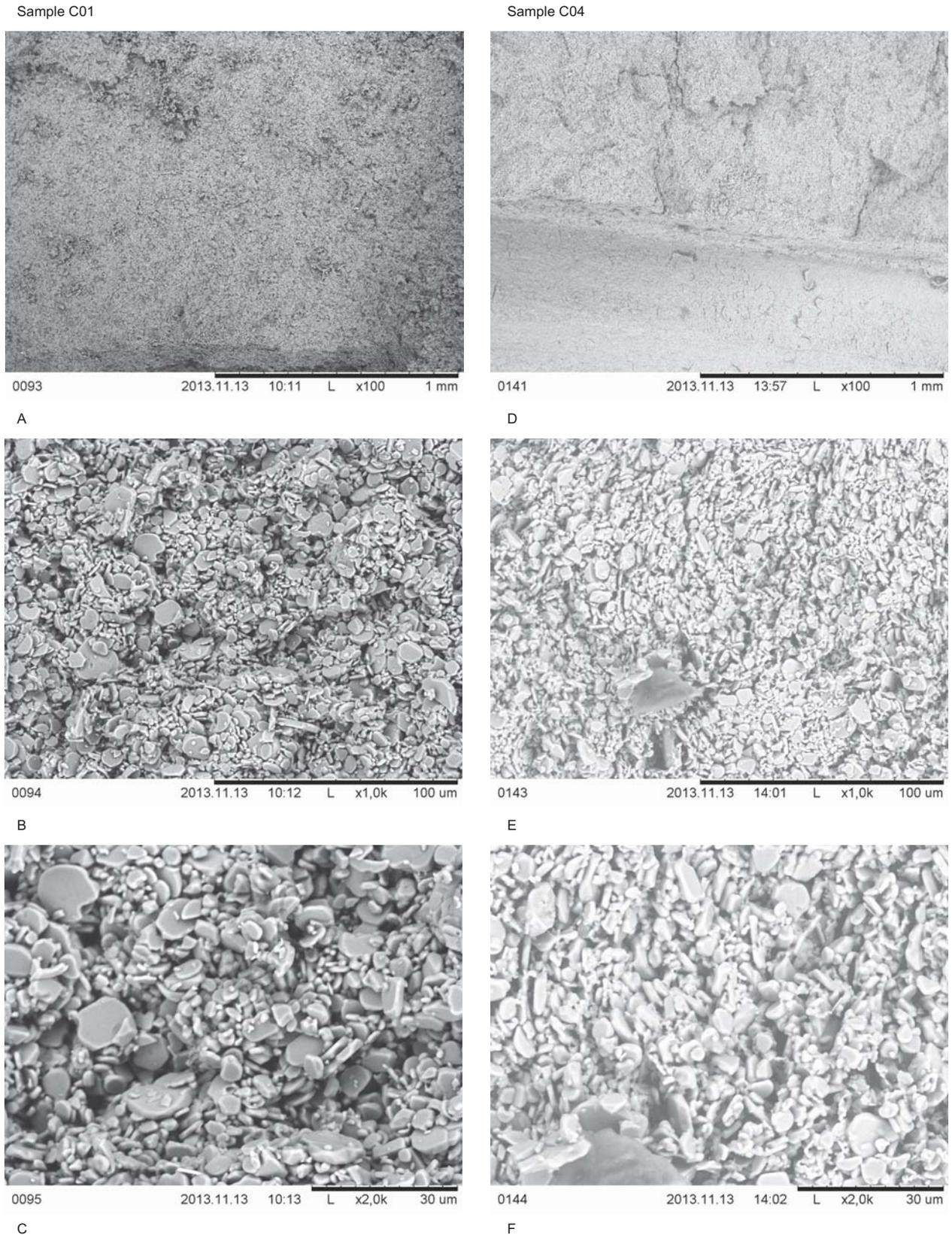

**Figure 11.** SEM micrographs of sintered C01 (left column, A–C) and C04 (right column, D–F) samples at various magnifications. Sample C01 contains alumina only, while C04 contains ATH too (see Table 1).

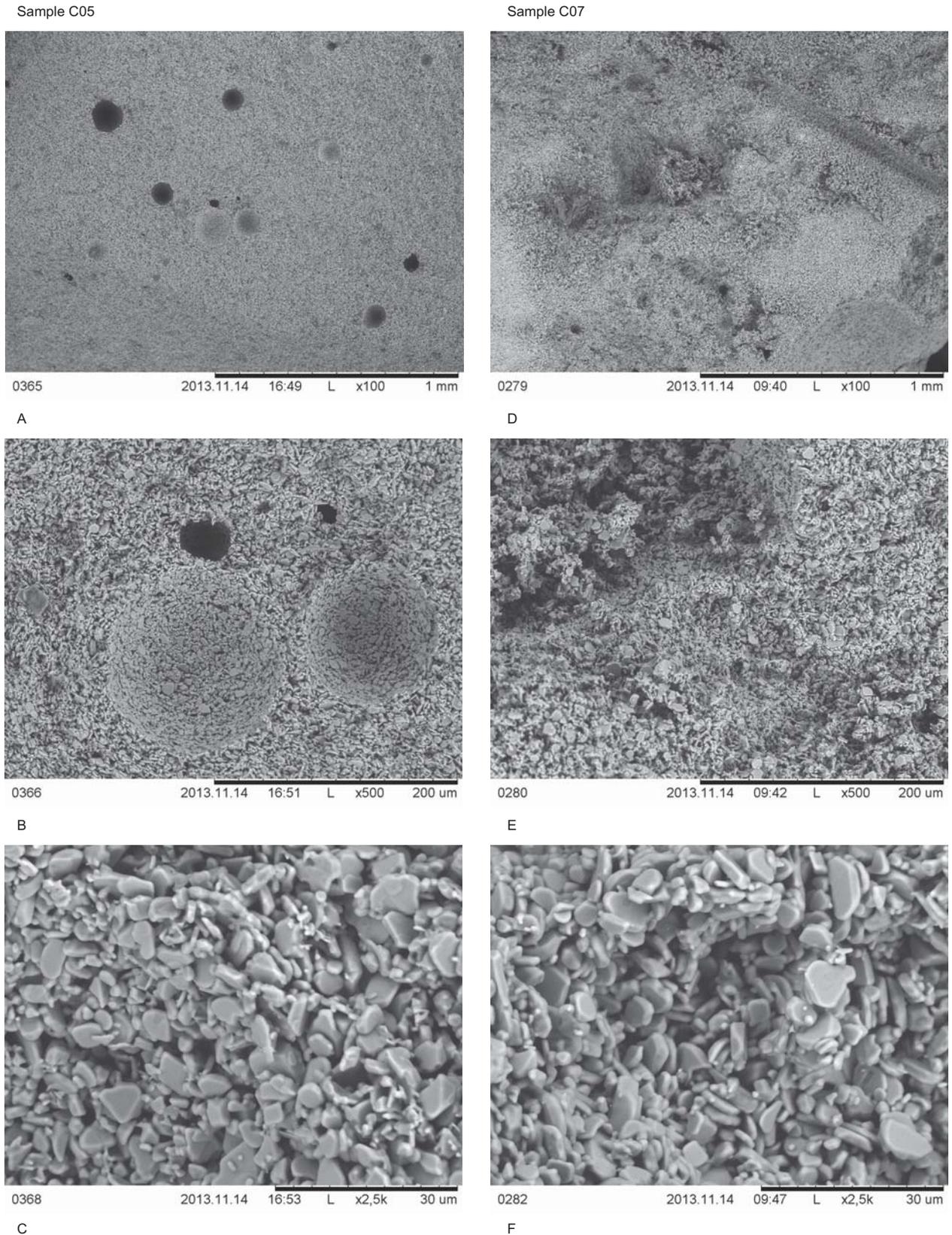

**Figure 12.** SEM micrographs of sintered C05 (left column, A–C) and C07 (right column, D–F) samples at various magnifications. Sample C05 contains ATH, while sample C07 citric acid as pore former. (See Table 1).

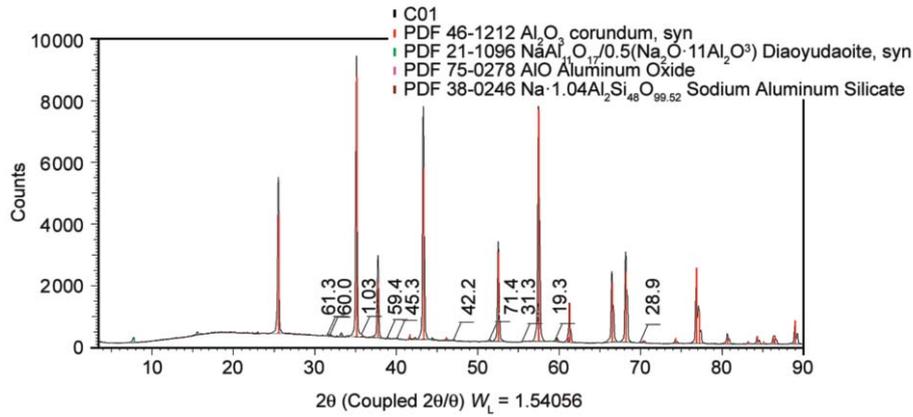

**Figure 13.** X-ray diffractogram of the sintered C01 sample. Diffraction lines of pure corundum are shown in red, other minor peaks in green.

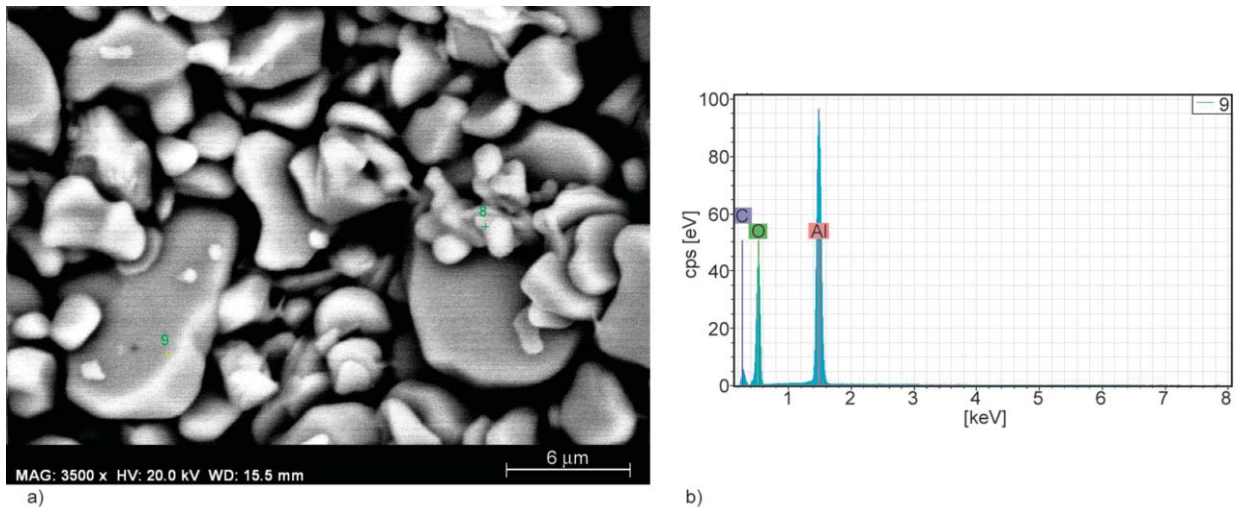

**Figure 14.** SEM/EDX micrograph of a sintered C01 sample. (Backscattered electron micrograph (a); EDX analysis at a single point (b). Disregarding a minor carbon peak the sample is pure alumina.

X-ray diffractogram shown in Figure 13 taken on the sintered sample of C01 shows that the materials consist almost purely of alumina (lines of corundum are shown for comparison). The same conclusion can be drawn from a SEM/EDX measurement (see Figure 14).

It can be concluded that samples C03, C05 and C06 containing UHMWPE, ATH and citric acid pore formers respectively exhibited higher porosity than the reference samples – although other compositions containing less ATH or citric acid alone did not. Other researchers report on micro-injected CIM/alumina parts with a relative density of 99.8% [47], others achieved similar relative densities with room temperature aqueous injection molding based on poly (vinyl pyrrolidone) solutions [48]. In our own aqueous slurry based low pressure molding experiment using ethylene-acrylic acid binder [19] the residual porosity was around 15%, i.e. similar to those values obtained in our CIM experiments, and the most probable pore size was around 1–3 μm. Further optimization of ceramic particle granulometry, composition (amounts of pore forming additives), de-binding and sintering schedule is needed to achieve a repeatable and predictable pore distribution in the CIM based products.

### 3.3. New approach to prepare oxide dispersed steel

In this case we tried to explore the possibility of adding finely distributed oxide particles (alumina, titania, silica using Al-stearate or KA322, Ti(iPrO)$_3$ or KR® TTS and TEOS respectively, see Table 2 in the experimental part) to stainless steel powder

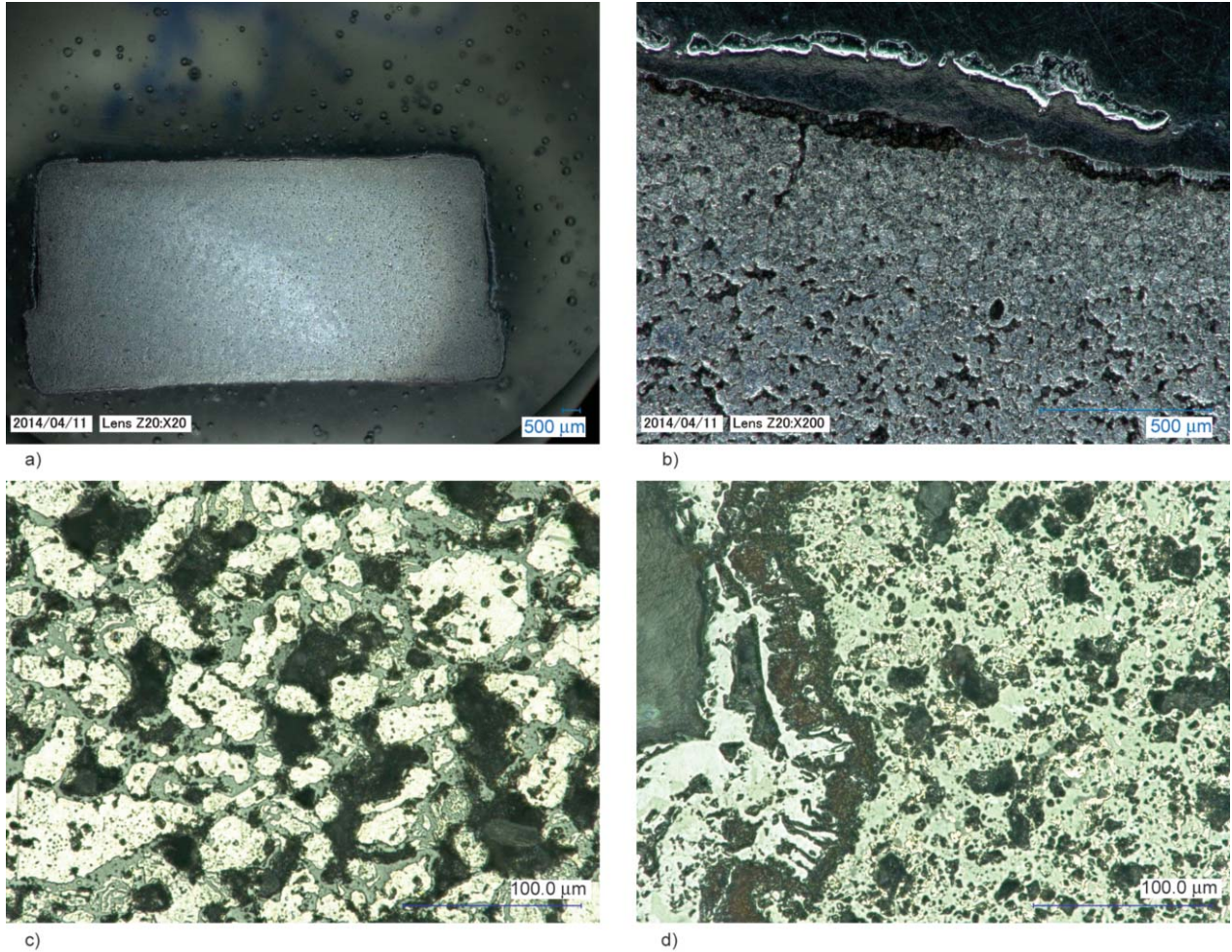

**Figure 15.** Optical micrographs of the sintered M01 sample, whole cross-section (a), the skin-core structure (b), higher magnification in the core region (c), higher magnification in the skin region (d). The core region seems to exhibit higher porosity

processed by MIM (actually, in our case by compression molding of MIM feedstock samples). The manufacturer of the stainless steel (SS) powder used by us, Höganäs suggests the use various lubricants, among others metal soaps, as Zn stearate as additives in powder metallurgy [49] but the authors do not discuss the appearance of oxide particles.

Densities of the sintered samples M01 and M02 were determined by the buoyancy method to establish the optimum sintering temperature, the calculated porosities are listed in Table 10. It is clear that simply by increasing the sintering temperature it is not possible to increase the density further – the porosity remains high (even if compared to the relatively high porosity of 18.8% quoted by the manufacturer of the SS powder for a 600 MPa compacted powder with a wax lubricant, see the Experimental part). It is clear that for structural parts (which are the typical applications of ODS steel [50]) high pressure consolidation methods as hot isostatic pressing or hot extrusion are needed [51] although high densities were also achieved by high voltage electric discharge compaction too [52]. As we had no direct access to these high temperature/high pressure processing technologies, we limited ourselves to morphological studies after ambient pressure sintering of the MIM feedstock samples after compression molding and de-binding.

The morphology and elemental composition distribution of samples listed in Table 2 were studied by optical microscopy and SEM/EDX respectively on samples cut from the sintered bodies described in the Experimental part and polished afterwards. Figure 15 shows the optical micrographs of sample M01 at the sintering temperature where its density was the highest at different magnifications. The presence of a core-skin structure is well discernible. There is a loosely bound skin layer but apparently the porosity is lower close to the periphery of the sample. No figures will be shown for sample M02, as they are very similar to M01. Nevertheless we chose the matrix composition of M01 for the further experiments, as we hoped that the presence of a copolymer containing polar co-monomers will help the uniform distribution of the metal-organic additives which are also moderately polar. Figure 16 shows, SEM/EDX micrographs at much higher magnification for sample M02 to demonstrate that the distribution of the alloying elements changes from point to point. Figures 16e and 16f were added (at a lower magnification) to show that the insignificant amount of Si alloying component is still detectable and is concentrated in certain regions. (It makes the evaluation of the TEOS-containing sample (M04) more complicated). Figure 17 shows the optical micrographs of the Ti-containing samples (M03 with titanium isopropoxide and M06 the KR® TTS titanate coupling agent with long fatty acid chains). In both cases both yellowish and pink inclusions can be seen, with sizes in the order of 10–50 μm. These inclusions are clearly non-metallic, it can be assumed that they are oxide glasses or microcrystals. The pink color may be attributed to the partial appearance of $Ti^{3+}$ ions in strongly reductive environments instead of the normal $Ti^{4+}$ ions. The yellowish color in other cases may be due to other contaminants. Local accumulations of Ti in M03 and M06 could be detected by SEM/EDX maps too (see Figure 18). Ti-rich areas appear at locations where the SEM micrograph is dark. The optical micrographs of the Al containing samples (M05 based on Al-stearate and M07 based on a more complex Al-coupling agent) are shown in Figure 19. Local accumulation of Al can be well detected by SEM/EDX maps of these samples (see Figure 20) – again in the "dark" areas (non-metallic compounds reflect the electrons less effectively). Figures 20c and 20d show that Al can be well detected point-wise too. Due to the presence of the Si alloying element in the SS powder used the detection of the local concentration of Si (presumably in amorphous $SiO_2$ form) is not as easy in M04 as it was in the case of other elements, but it is possible. The optical micrographs (Figures 21a and 21b), a comparison of the element distribution in bright and "dark" areas of the SEM micrograph (Figures 21c–21f) and the Si distribution on a SEM/EDX micrograph exhibiting an Si-rich are at a "dark" spot (Figures 21g and 21h) clearly demonstrate the presence of Si-containing inclusions in the sintered SS sample. The green color of the silicate inclusions on the optical micrographs is presumably due to the presence of $Fe^{2+}$ ions in the glassy structure.

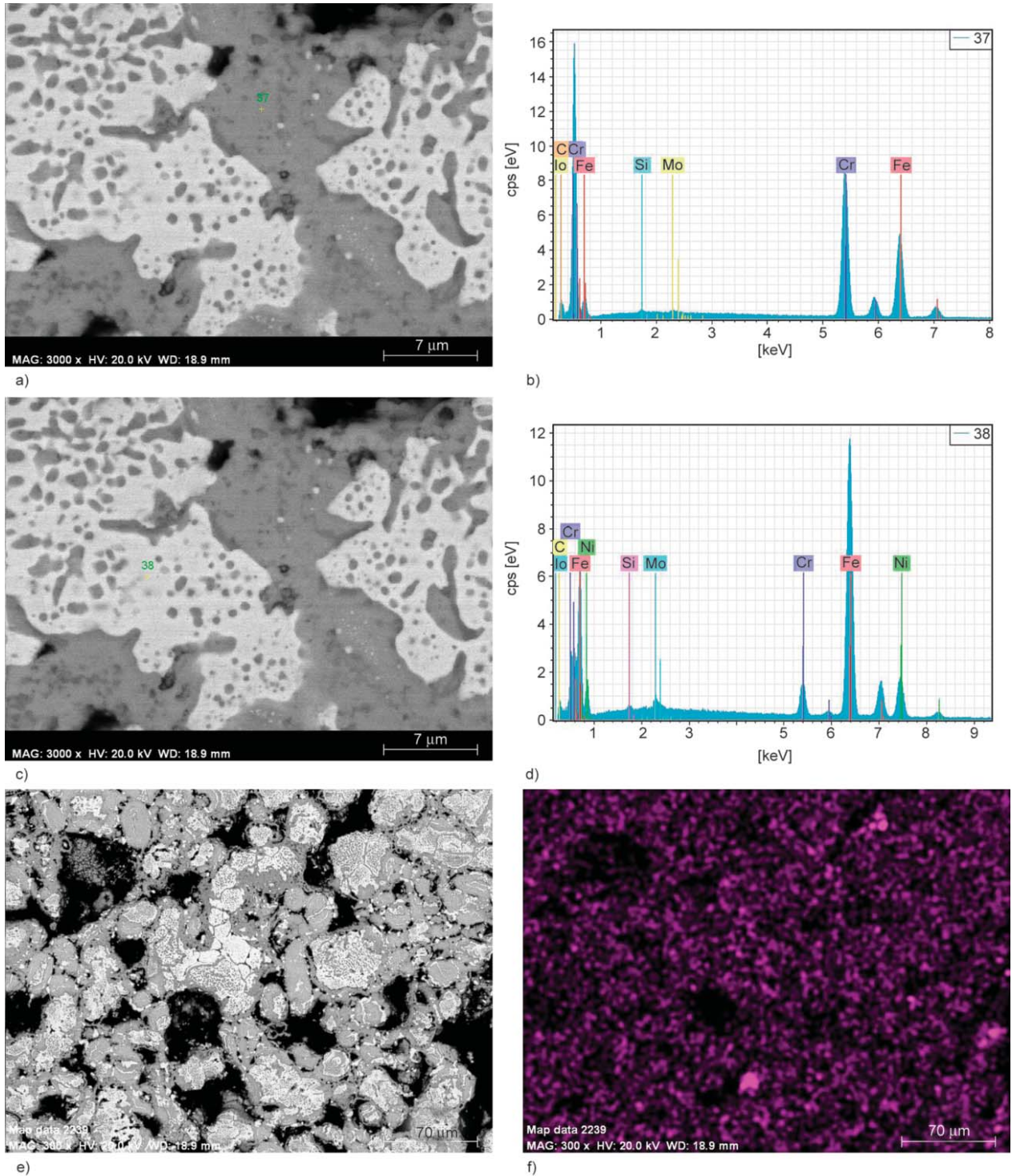

**Figure 16.** SEM micrographs of the sintered M02 sample, backscattered electron micrograph and EDX analysis at a single point in a darker area (a and b) – more alloying elements; backscattered micrograph and EDX analysis at a single point in a brighter area (c and d) – more iron; backscattered micrograph and the distribution of the Si alloying element (e and f).

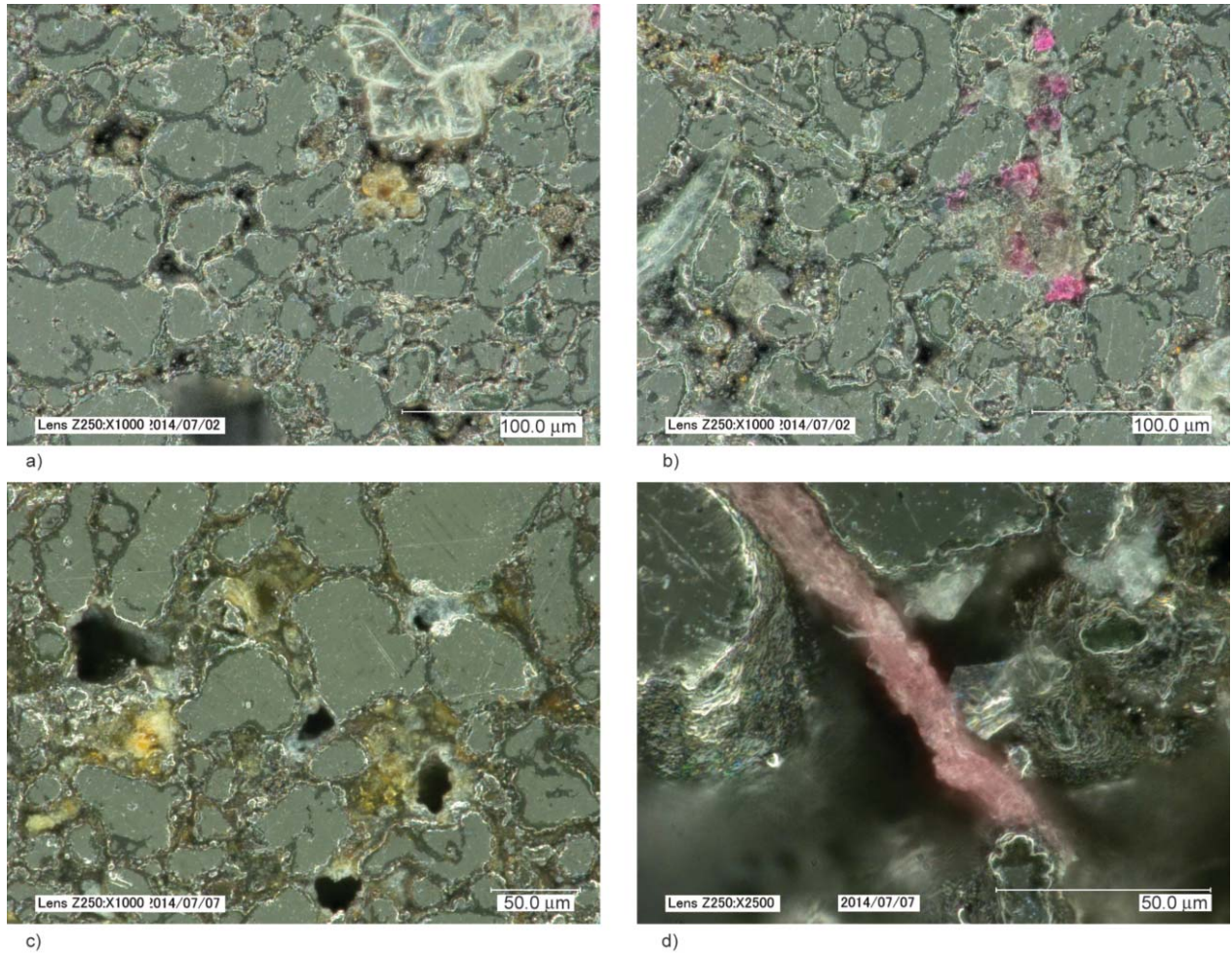

**Figure 17.** Optical micrographs of the sintered M03 (a, b) and M06 (c, d) samples containing Ti(iPrO)$_4$ and KR® TTS additive respectively. In both samples non-metallic inclusions can b detected both with pale yellow and with pink colors.

**Table 10.** Experimentally determined densities of M01 and M02 (presented in Table 2) samples after various sintering schedules together with the calculated porosities (using 8.00 g/cm$^3$ for the pure metal component).

| Sample | Sintering temperature [°C] | Density [g/cm$^3$] | Porosity [%] |
|---|---|---|---|
| M01 | 950 | 5.00 | 37.5 |
| M01 | 1050 | 5.30 | 33.8 |
| M01 | 1150 | 5.47 | 31.6 |
| M01 | 1300 | 4.65 | 41.9 |
| M02 | 950 | 5.24 | 34.5 |
| M02 | 1050 | 5.40 | 32.5 |
| M02 | 1150 | 5.06 | 36.8 |
| M02 | 1300 | 4.56 | 43.0 |

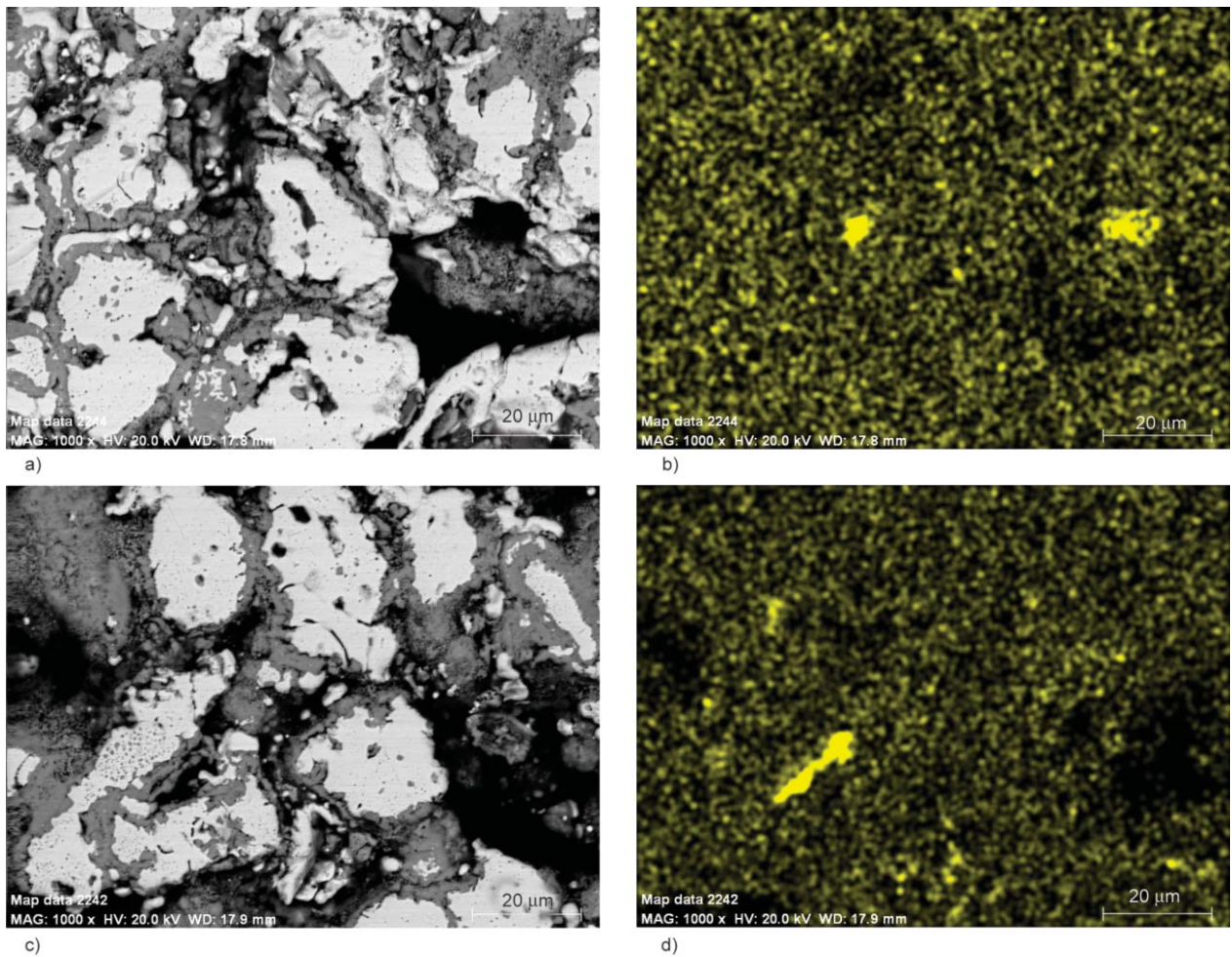

**Figure 18.** Backscattered electron micrographs and the distribution of Ti element in sintered M03 (a, b) and M06 (c, d) samples. In both samples local concentration of Ti can be detected in "dark" some areas of the backscattered SEM micrograph.

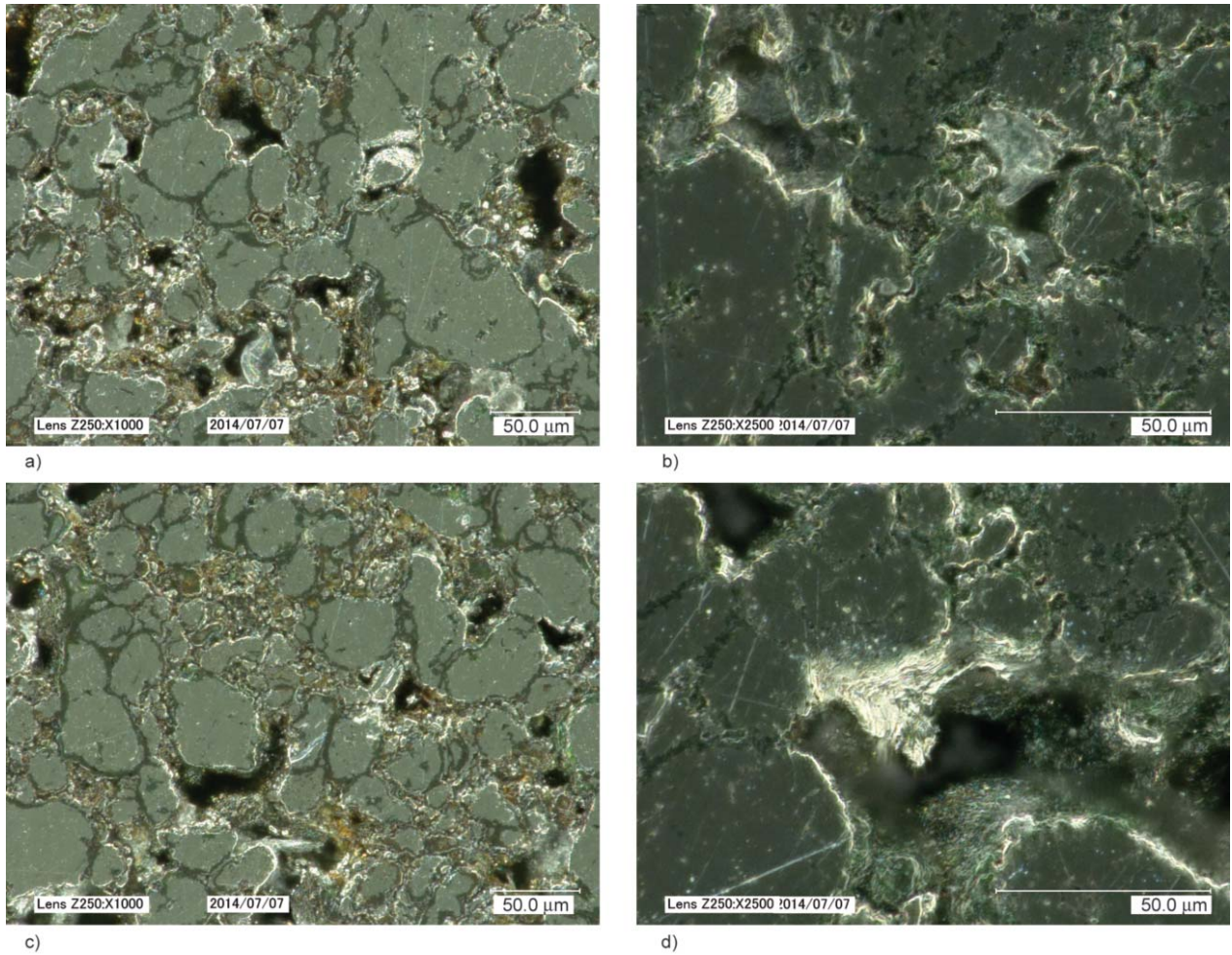

**Figure 19.** Optical micrographs of the sintered M05 (a, b) and M07 (c, d) samples containing Al-stearate and KA 322 additives respectively. Non-metallic inclusions can be detected.

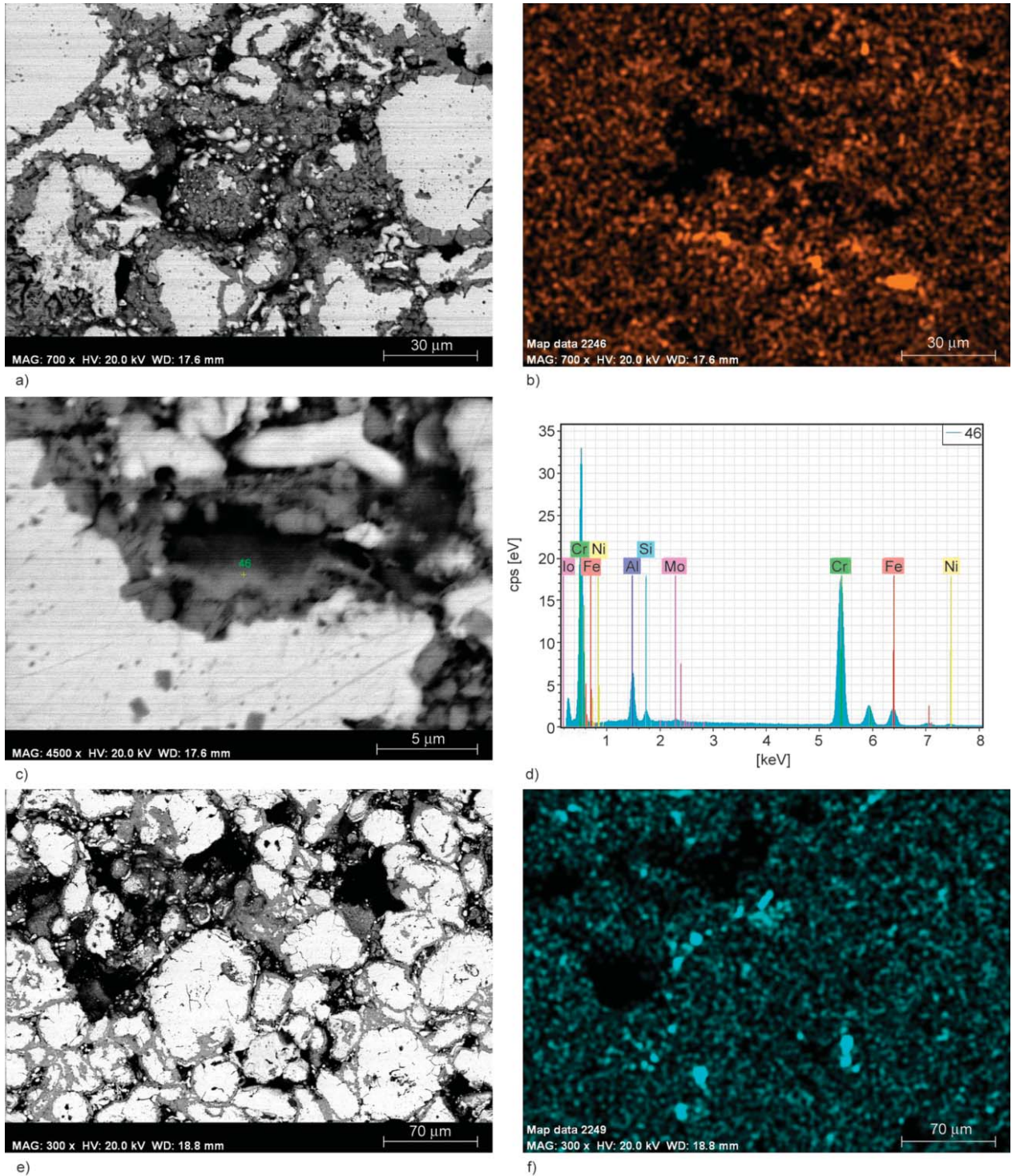

**Figure 20.** Backscattered electron micrographs and the distribution of Al element in sintered M05 (a, b), backscattered electron micrograph and EDX analysis at a "dark" spot of M05 (c, d); backscattered electron micrograph and the distribution of Al element in sintered M07 (e, f) samples. Aluminum appears together with other alloying elements. In both samples local concentrations of aluminum can be observed.

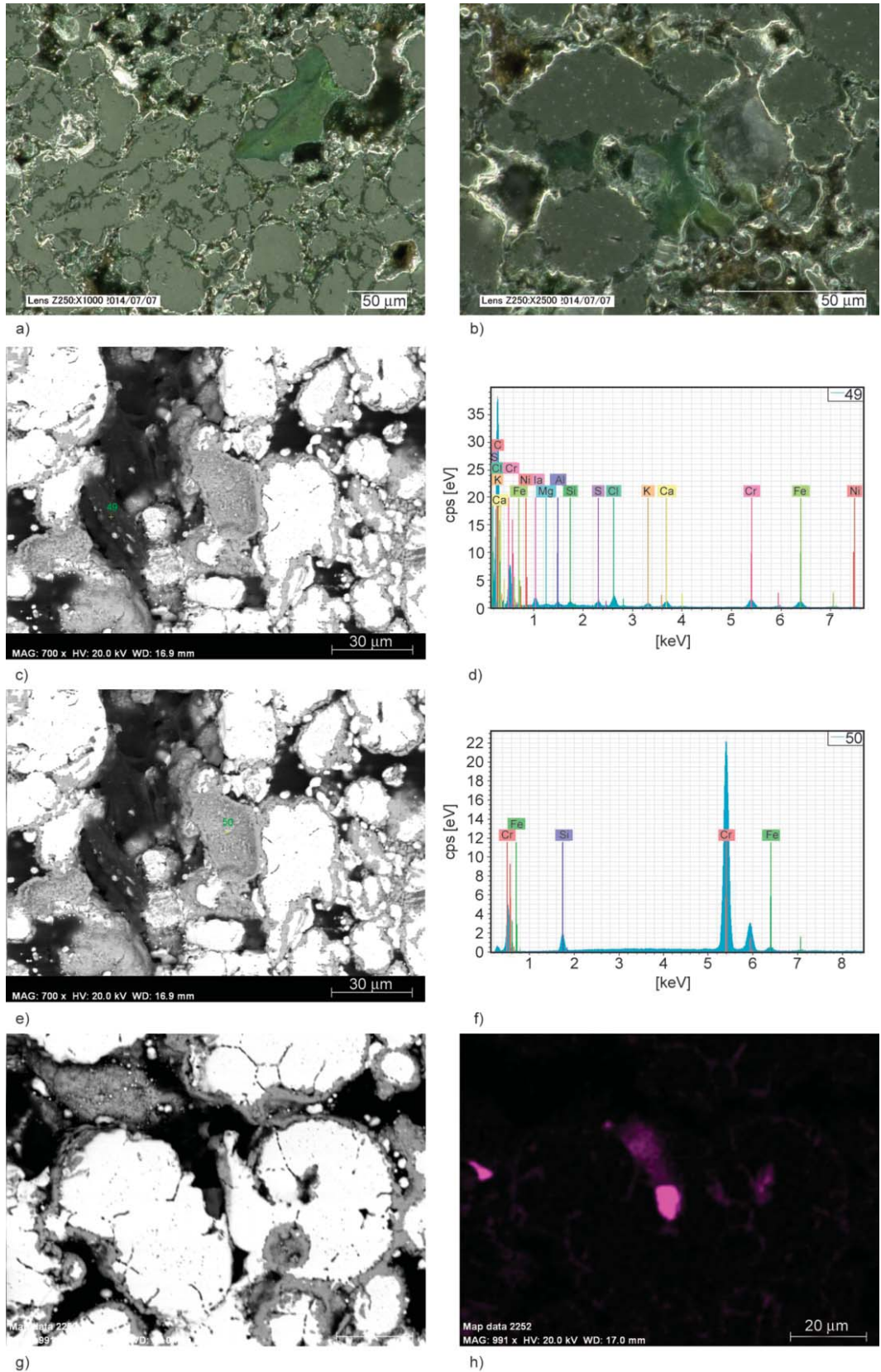

**Figure 21.** Optical micrographs of the sintered M04 (a, b) sample; backscattered electron micrograph and EDX analysis at a "dark" spot of M04 (c, d) (Si occurs together with C and several other contaminants); backscattered electron micrograph and EDX analysis at a "dark" spot of M04 (e, f) (Si occurs together with iron and other SS alloying elements); Concentration of Si at a "dark spot" of M04 (g, h). The optical micrographs show pale green non-metallic inclusions. Si may appear together with alloying elements – this is probably metallic Si, while it appears also in separated form, probably in silica inclusions.

In spite of the fact that these experiments gave half-success at best, as the porosity of the samples was too high to achieve high strength (sintered samples M03, M04 were relatively homogeneous and strong while samples M05, M06 and M07 broke more easily, exhibited inhomogeneity and porosity) and the size distribution of the inclusions was far from the nanometric scale necessary for the reinforcing effects observed in ODS steels, we could clearly demonstrate that by using metal-organic compounds it is possible to admix metal, semi-metal and metal oxide particles to stainless steel powder. We analyzed the potential reasons of partial failure to be able to give recommendations to improve the process. One reason is that during the de-binding process when the organic components degrade and the metal-organic components are transformed into the corresponding oxides, the organic phase begins to flow and due to the surface tension differences the organic component (together with the metal-organic compounds) tends to accumulate at the interstices between the SS powder particles – and this distribution is fixed during the sintering process. This problem presumably could be reduced if the MIM process is replaced by a combination of the proposed method and mechanical alloying: if one ball-mills SS powder with a calculated amount of the selected metal-organic compound and uses a liquid as processing aid which at least partially dissolves the metal-organic compound. This should result in a more homogeneous distribution of the additive and presumably some chemical reactions would also occur at the surface of the metal particles. A mild hydrolysis of the metal-organic compounds (e.g. by vapors) could possibly anchor the oxide precursor on the steel surface – thus reducing the tendency to accumulate in the interstices. If this pre-treated powder is exposed to high pressure compression or extrusion accompanied by gradual heating it is conceivable that alloys with better properties are obtained. If these modifications prove useful, a wide range of possibilities opens, as almost all metallic, semi-metallic or non-metallic elements have a wide variety of metal-organic derivatives which can be optimized for this application.

## 4. Conclusions

In the three examples we tried to demonstrate the extreme variability of hybrid composites systems and the use of up-to-date analytical techniques to understand those chemical and phase transformations which are frequently accompanied by the partial or complete degradation or elimination of the organic components. These organic components nevertheless play a major role in determining the final composition and morphology of the final composite products obtained. Their rheology, polarity, chemical activity, adhesion are important factors which have to be taken into taken into account when designing the process. In the heat resistant prepreg the organic matrix is a structural part of the lining system and becomes a "sacrificial" component only if it is exposed to extreme heat. In the examples related to CIM and MIM feedstocks transformed into sintered bodies two kinds of organic additives have to be considered: the polymer and wax components needed to embed the ceramic or metal particles while the feedstock composition is being mixed and processed into the final shape and other additives used to modify the final composition. In the case of the CIM material there were (organic and inorganic) additives which give rise to gases that produce porosity, while in the case of the new MIM material metal-organic additives were used in order to be miscible (or at least technically compatible) with the organic matrix but able to transform into metal-oxide particles during de-binding and sintering. Of course, this distinction is not very sharp, as these organic components interact with each other and the inorganic components. Thermoanalytical methods, in combination with surface analytical techniques and various microscopic methods help to understand changes in chemical composition and morphology. TGA and DTA are the basic tools to identify the phase transitions (melting and evaporation) and chemical reactions, degradation of the component alone and mixed together. Vibrational spectroscopic methods help to understand the chemical transformation of the organic components until they are not completely eliminated at higher temperatures. XPS provides detailed and sensitive information on the chemical composition of the topmost layers, while SEM/EDX gives information about both morphology and element distribution in somewhat deeper layers. Even optical microscopy proved to be useful aid at lower magnifications, especially in understanding the MIM transformation process and in identifying the non-metallic components obtained.

These examples hopefully give an insight into the exciting world of hybrid composites.

## Acknowledgement

This paper is based on an invited lecture presented at the 2nd International Conference on Science and Engineering of Materials, Sharda University, Greater Noida, India, January 2018.

Thanks are due to the companies (mentioned in the text) providing material or product samples, especially to Dr. Gábor Nagy, CEO of Polinvent Ltd. not only or the samples but for the useful discussions. The authors thank Aladár Szabó for sharing the results of his MSc, which are used in the discussion. Thanks are due to Dr. Attila Farkas for the Raman, Dr. József Hári for the FTIR, Balázs Pinke for the SEM/EDX tests measurements.

## References

[1] Y. Swolfs, L. Gorbatikh, I. Verpoest: Composites Part A 67 (2014) 181.
https://doi.org/10.1016/j.compositesa.2014.08.027

[2] V. K. Thakur, M. K. Thakur, A. Pappu: Hybrid Polymer Composite Materials, 1st Ed. Properties and Characterisation, Woodhead Publishing (2017).

[3] M. Guglielmi, G. Kickelbick, A. Martucci (Eds.), Sol-Gel Nanocomposites, Springer, Berlin, (2014).

[4] J, González-Gutiérrez, G. Beulke Stringari, I. Emri: Powder Injection Molding of Metal and Ceramic Parts, in: Jian Wang (Ed.), Some Critical Issues for Injection Molding, ISBN: 978-953-51-0297-7, InTech, Available from: http://www.intechopen.com/books/some-critical-issues-for-injection-molding/powderinjection-molding-of-metal-and-ceramic-parts (2012) 65.

[5] E. Bernardo, L. Fiocco, G. Parcianello, Enrico Storti, P. Colombo: Materials 7 (2014) 1927.
https://doi.org/10.3390/ma7031927

[6] M. Szycher: Szycher's Handbook of Polyurethanes (2nd Ed.), Taylor and Francis Group, London, (2013).

[7] Polinvent Kft: Process for Producing Filled Systems Based on Polyisocyanate/Polysilicic Acid of Special Structure, Hungarian Patent: HU212440 (B) (1996).

[8] J. Bodi, Z. Bodi, J. Scucka, P. Martinec: Polyurethane Grouting Technologies, in: F. Zafar (Ed.), Polyurethane InTech,. Available from: https://www.intechopen.com/books/polyurethane/polyurethane-grouting-technologies (2012).
https://doi.org/10.5772/35791

[9] J. L. Provis, S.A. Bernal: Annu. Rev. Mat. Sci. 44 (2014) 299.
https://doi.org/10.1146/annurev-matsci-070813-113515

[10] N. Castella, S. Grishchuk, J. Karger-Kocsis, M. Unik: J. Mater. Sci. 45 (2010) 1734.
https://doi.org/10.1007/s10853-009-4145-9

[11] M. An, D. Shi, G. Wang: Int. J. of Chemistry 1 (2009) 10.
https://doi.org/10.5539/ijc.v1n2p10

[ 2] Polinvent Kft.: Polyisocyanate and waterglass based hybrid resins, composites containing them and process to produce them, EP1791886 (B1) (2008).

[13] Polinvent Kft.: Fibre reinforced prepreg and its applications, EP2826811 (A2) (2015).

[14] D. Bleyan, B. Hausnerova: Key Eng. Mater. 581 (2014) 82.
https://doi.org/10.4028/www.scientific.net/KEM.581.82

[15] P. Rogers, R. Jain: Mater. Manuf. Processes (2014).
https://doi.org/10.1080/10426914.2014.984214

[16] Zs. Karácsony, A. Erős, E. Andersen, Gy. Bánhegyi: Mater. Sci. Forum 812 (2015) 95.
https://doi.org/10.4028/www.scientific.net/MSF.812.95

[17] Á. Egész, L. A. Gömze: építőanyag 65 (2013) 107.
https://doi.org/10.14382/epitoanyag-jsbcm.2013.20

[18] A. Szabó: MSc Thesis, Preparation and investigaton of injection molding feedstocks with polymer matrix (in Hungarian), Miskolc University, Hungary, (2014).

[19] L. Angyal, Zs. Mátyás-Karácsony, A. Kállay-Menyhárd, Gy. Bánhegyi: IOP Conf. Series Mater. Sci. Eng. 123 (2016) 012039.
https://doi.org/10.1088/1757-899X/123/1/012039

[20] G. Matula, J. Krzysteczko: Journal of Achievements in Materials and Manufacturing Engineering 71 (2015) 14.
https://doi.org/ –

[21] W. Liu, Z. Xie: Science of Sintering 46 (2014) 3.
https://doi.org/10.2298/SOS1401003L

[22] J. A. Reglero Ruiz, M. Vincent, J.-F. Agassant, T. Sadik, C. Pillon, Ch. Carrot: Polym. Eng. Sci. 55 (2015) 2018.
https://doi.org/10.1002/pen.24044

[23] W. Hzez: MSc Thesis, Processing and Properties of Porous Alumina Ceramics by Gel Casting, Luleå University of Technology, Sweden, (2015).

[24] J.S. Magdeski: Journal of the University of Chemical Technology and Metallurgy, 45 (2010) 143.
https://doi.org/ – , ISSN 1314-3859

[25] Y. B. P. Kwan, D.J. Stephenson, J.R. Alcock: J. Mater. Sci., 35 (2000) 1205.
https://doi.org/10.1023/A:1004792605528

[26] A. Sarikaya, F.Dogan: Ceram. Int. 39 (2013) 403.
https://doi.org/10.1016/j.ceramint.2012.06.041

[27] L. Raman, K. Gothandapani, B.S. Murty: Defence Sci. Journal, 66 (2016) 316.
https://doi.org/10.14429/dsj.66.10205

[28] S. Kavithaa, R.Subramanian, P.C. Angelo: Trans. Indian Inst. Met. 63 (2010) 67.
https://doi.org/10.1007/s12666-010-0010-4

[29] X. Boulnat, D. Fabregue, M. Perez, M-H. Mathon, Y. De Carlan: Metall. Mater. Trans. A 44 (2013).
https://doi.org/10.1007/s11661-013-1719-6

[30] Q.X. Sun, T. Zhang, X.P. Wang, Q.F. Fang, T. Hao, C.S. Liu: J. Nucl. Mater. 424 (2012) 279.
https://doi.org/10.1016/j.jnucmat.2011.12.020

[31] F. Bergner, I. Hilger, J. Virta, J. Lagerbom, G. Gerbeth, S. Connolly, Z. Hong, P. S. Grant, T. Weissgärber: Metall and Mat. Trans. A, 47A (2016) online.
https://doi.org/10.1007/s11661-016-3616-2

[32] K. Lindqvist, E. Carlström, M. Persson, R. Carlsson: J. Am. Ceram. Soc. 72 (2005) 99.
https://doi.org/10.1111/j.1151-2916.1989.tb05960.x

[33] Silane Copuling Agents, Gelst, Inc., https://www.gelest.com/wp-content/uploads/Goods-PDF-brochures-couplingagents.pdf, donwloaded 2017.10.19.

[34] Ken-React® Titanate, Zirconate & Aluminate Coupling Agents & Catalysts, http://4kenrich.com/products-and-benefits/full-products-list/ken-react-titanate-zirconate-aluminate-coupling-agents-catalysts/, downloaded 2017.10.19.

[35] J. M. Jang, W. Lee, S.-H. Ko, C. Han, H. Choi: Arch. Metall. Mater. 60 (2015) 1281.
https://doi.org/10.1515/amm-2015-0114

[36] Product Data Sheet 316/316L Stainless Steel, AK Steel, http://www.aksteel.com/pdf/markets_products/stainless/austenitic/316_316l_data_sheet.pdf, download 2017.10.19.

[37] M. Chanda, S.K. Roy: Plastics Technology Handbook, 3rd ed., Marcel Dekker, New York, (1998), p. 186.

[38] M. Mohai: Surf. Interface Anal. 36 (2004) 828. https://doi.org/10.1002/sia.1775;
http://www.chemres.hu/aki/ XMQpages/XMQhome.htm

[39] S. Evans, R. G. Pritchard, J. M. Thomas: J. Electron Spectrosc. Relat. Phenom. 14 (1978) 341.
https://doi.org/10.1016/0368-2048(78)80008-5

[40] R. F. Reilman, A. Msezane, S. T. Manson: J. Electron Spectrosc. Relat. Phenom. 8 (1976) 389.
https://doi.org/10.1016/0368-2048(76)80025-4

[41] I. Prencipe, D. Dellasega, A. Zani, D. Rizzo, M. Passoni: Sci.Technol. Adv. Mater. 16 (2015) 025007.
https://doi.org/10.1088/1468-6996/16/2/025007

[42] K. Shimizu, Ch. Phanopoulos, R. Loenders, M.-L. Abel, J. F. Watts: Surf. Interface Anal. 42 (2010) 1432.
https://doi.org/10.1002/sia.3586

[43] H. Elderfield, J.D. Hem: Mineral. Mag. 39 (1973) 89.
https://doi.org/10.1180/minmag.1973.039.301.14

[44] M. Kayalvizhi, J. Suresh, S. Karthik, A. Arun: Int. J. Plast. Technol. 20 (2016) 128.
https://doi.org/10.1007/s12588-016-9145-4

[45] J. Hu, Z. Chen, Y. He, H. Huang, X. Zhang: Res. Chem. Intermed. 43 (2017) 2799.
https://doi.org/10.1007/s11164-016-2795-1

[46] G. A. Pitsevich, M. Shundalau, M. A. Ksenofontov, D. S. Umreiko: Global J. of Anal. Chem. 2 (2011) 114.
https://doi.org/ -

[47] P. Thomas, B. Levenfeld, A. Várez, A. Cervera: Int. J. Appl. Ceram. Technol. 8 (2011) 617.
https://doi.org/10.1111/j.1744-7402.2009.02471.x

[48] V.L. Wiesner, J.P. Youngblood, R.W. Trice: J. Eur. Ceram. Soc. 34 (2014) 453.
https://doi.org/10.1016/j.jeurceramsoc.2013.08.017

[49] M. Larsson, M. Ramstedt: Lubricants for compaction of P/M components, www.ipen.br/biblioteca/cd/ptech/2003/PDF/09_12.pdf, dowloaded 2017.10.19.

[50] C. Capdevila, M. Serrano, M. Campos: Materials Science and Technology, 30 (2014) 1655 (Online).
https://doi.org/10.1179/0267083614Z.000000000787

[51] M. De Sanctis, A. Fava, G. Lovicu, R. Montanari, M. Richetta, C. Testani, A. Varone: Metals 7 (2017) 283.
https://doi.org/10.3390/met7080283

[52] I. Bogachev, A. Yudin, E. Grigoryev, I. Chernov, M. Staltsov, O. Khasanov, E. Olevsky: Materials 8 (2015) 7342.
https://doi.org/10.3390/ma8115381